\documentclass[twocolumn]{aastex631} %

\usepackage{amsmath}
\usepackage{booktabs} 
\usepackage{comment}
\usepackage{graphicx}   
\usepackage{multirow}   
\usepackage{etoolbox}
\usepackage{float}

\begin{document}

\title{Stellar Evolution with Radiative Feedback in AGN Disks}

\shorttitle{AGN Stars with Radiative Feedback}
\shortauthors{Xu et al.} 

\author{Zheng-Hao Xu}
\affiliation{Department of Physics, University of California Santa Barbara, Santa Barbara, CA, 93106, USA}

\author{Yi-Xian Chen}
\affiliation{Department of Astrophysics, Princeton University, Princeton, NJ, 08544, USA}

\author{Douglas N. C. Lin}
\affiliation{Department of Astronomy and Astrophysics, University of California, Santa Cruz, CA, 95064, USA}
\affiliation{Institute for Advanced Study, Tsinghua University, Beijing, 100084, China}
\affiliation{Department of Astronomy, Westlake University, Hangzhou, Zhejiang, 310030, China}

\begin{abstract}
Stars embedded in the {inner} pc region of an active galactic nucleus (AGN) experience extreme accretion conditions that significantly alter their evolution. 
We present one-dimensional MESA simulations of stars growing and decaying within AGN disks, implementing radiative-feedback-regulated accretion which limits stellar growth near the Eddington luminosity, as well as wind-driven mass loss. 
Unlike stand-alone stars in the field, these 
embedded stars follow unique evolutionary tracks with well-determined mass evolution and chemical yields. 
We distinguish two regimes: ``immortal" stars that indefinitely remain on the main sequence due to efficient hydrogen mixing; and ``metamorphic" stars that evolves off the main sequence, ultimately enriching the disk with heavy elements upon hydrogen and helium exhaustion in their cores. 
Results indicate that embedded stars in AGN disks can attain large masses, but gas retention and limited mixing likely render the ``immortal" track unsustainable. 
{We show radiative feedback plays a critical role in preventing runaway growth, 
since it regulates the inflow to at most of 
order-unity the Eddington-limited mass-loss rate.}
Embedded metamorphic stars significantly enrich AGN disks with helium and $\alpha$-elements, 
potentially explaining the observed high metallicity in broad-line regions (BLR) without excessive helium enrichment. 
This study underscores the critical interplay between stellar feedback and accretion physics in shaping the stellar populations and chemical evolution within AGN disks.
\end{abstract}

\keywords{ active galactic nuclei --- stellar evolution --- radiative feedback --- mass loss --- chemical enrichment}

\section{Introduction}

Accretion disks in active galactic nuclei (AGNs) provide environments with extreme density and radiation fields.
In the dense gas of an AGN disk, stars can form in situ via disk fragmentation \citep{GoodmanTan2004,JiangGoodman2011,Chen2023} or be captured from the nuclear star cluster \citep{Artymowicz1993,MacLeodLin2020,Wang2024}.
Once formed or embedded, 
these stars are immersed in gas with densities $\rho \sim 10^{-20}$–$10^{-10}$ g cm$^{-3}$ and temperatures $T \sim 10^3$–$10^6$ K, 
which are orders of magnitude higher than those in typical interstellar environments \citep{Cantiello2021}. 
Their energy output may help sustain the extended, self-gravitating disk structure \citep{SirkoGoodman2003,Thompson2005,ChenLin2024}, potentially relevant to the infrared emission in `Little Red Dots' \citep{Zhang2025}. 

Several lines of evidence support the survival and growth of stars within AGN disks. Emission lines from the broad-line region (BLR) consistently show redshift-independent, super-solar metallicities \citep{Hamann1999,Hamann2002,Nagao2006,Xu2018,Wang+2022,Huang2023}, possibly due to enrichment from local stellar populations \citep{AlidibLin2023,Fryer2025}. 
Stars in AGN disks may also serve as progenitors of gravitational wave sources, 
either through stellar mergers or the formation of massive black hole binaries within the disk \citep{McKernan2012,graham2020,Tagawa2020a,Samsing2022}, 
when direct capture of compact objects is less efficient \citep{MacLeodLin2020}. 

Regarding these scenarios, 
the extent to which massive stars contribute to the metallicity of AGN disks and evolve off the main sequence depends sensitively 
on their poorly understood evolution in the gas-rich AGN environment. 
\citet{Jermyn2022} emphasized that such stars might become ``immortal" if the hydrogen-rich accreted gas can efficiently diffuse through the radiative zone, such that they continuously burn hydrogen over the AGN’s lifetime.
These immortal stars could steadily burn hydrogen near the Eddington limit, 
influencing the disk’s chemical composition by converting hydrogen into helium (and potentially releasing it back into the disk). 
However, this raises a critical question: 
while AGN broad-line regions exhibit enhanced metallicities, particularly in $\alpha$-elements and iron, 
they do not show the extreme helium enrichment \citep{Huang2023}
that would be expected if immortal stars were common. 

More broadly, using accretion and stellar wind prescriptions within one-dimensional stellar evolution models, 
\citet{Dittmann2021, Fabj2025} conducted parameter surveys across typical AGN disk environments and found that stellar evolution outcomes vary widely depending on the local gas density. 
In high-density regions, 
stars can not only become fully immortal, 
but also their Kelvin–Helmholtz timescale may fall below the mass doubling timescale, triggering runaway growth toward pair-instability supernovae and/or intermediate-mass Black Holes. 
Even a small number of such supermassive stars could dramatically alter disk properties—making it difficult to reconcile their existence with the lack of corresponding observational signatures from AGNs.

However, a key physical ingredient missing from these studies is a self-consistent treatment of radiative feedback. 
Recent hydrodynamic simulations in \citet{Chen2024, Chen2025} demonstrate that when full radiative transfer is included, 
the accretion rate onto stars is limited by Eddington feedback.  
This significantly 
reduces the stars' actual accretion rate compared to the Bondi rate usually assumed in 1D studies, 
placing a fundamental limit on stellar growth in AGN disks, similar to the Eddington limit for black holes.

In this work, we present the Stellar Evolution and Pollution in AGN Disks (SEPAD) model that incorporates
an updated treatment of radiative feedback. 
{This mechanism plays a critical role in preventing runaway growth in high-density environments, 
since radiative feedback self-regulates the inflow to at most of order-unity the Eddington-limited mass-loss rate.}
We emphasize that once stars reach high luminosities, 
their accretion becomes limited by the Eddington rate, 
making the subsequent evolution effectively \textit{independent} of the background gas density.

The paper is organized as follows:
In §\ref{sec:theory}, 
we develop a theoretical framework for accretion onto stars embedded in AGN disks, 
including Bondi accretion estimates, 
radiative feedback limits, and semi-analytical estimates for stellar quasi-equilibrium masses.
In §\ref{sec:methods}, 
we describe our numerical approach using MESA, detail the modifications made to the accretion and wind routines from \citet{Cantiello2021}, 
and explain our implementation of mixing processes within the radiative zones.
In §\ref{sec:res}, 
we present results from representative simulations, 
examining stellar mass growth across different conditions, 
the interplay between accretion and wind mass loss, 
average luminosity and Eddington ratios, 
surface composition evolution, and the final stellar outcomes.
Finally, 
in §\ref{sec:sum}, we summarize our key conclusions, 
discuss the broader implications of our findings for observations, 
including their relevance to BLR emission properties and 
the overall energy budget of AGN disks.and suggest directions for future work.

\section{Analytical Expectations}\label{sec:theory}

\begin{figure*}[htbp]
  \centering
  \includegraphics[width=0.7\textwidth]{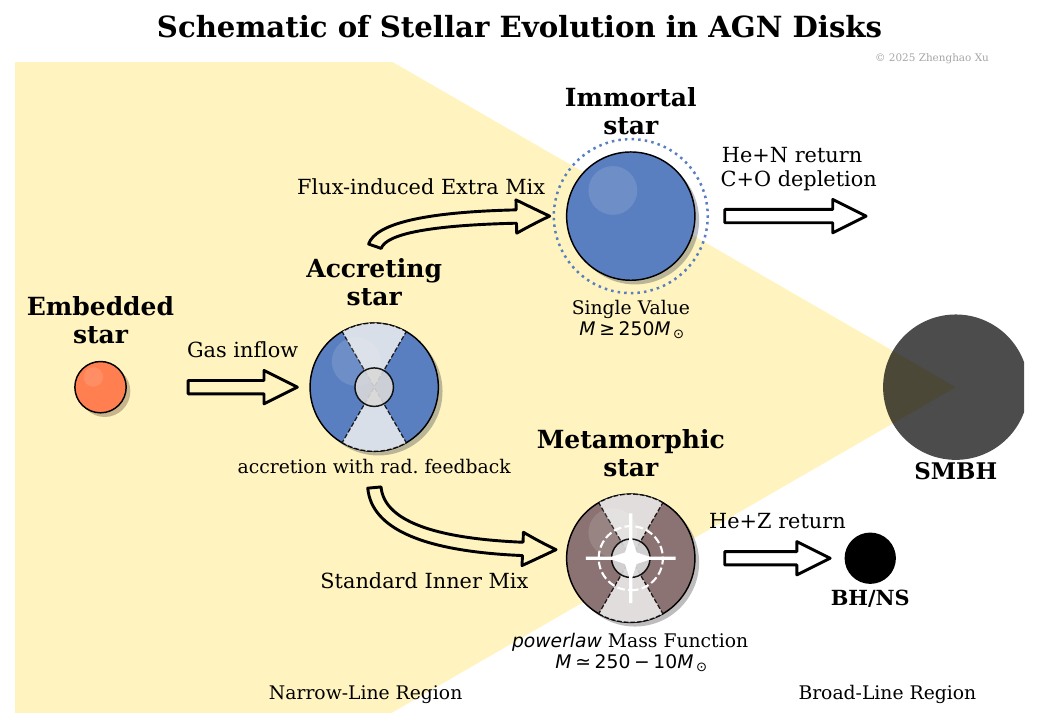}
  \caption{ Schematic of a star embedded in an AGN disk. 
Disk gas feeds the star at a Bondi–Hoyle rate but radiative feedback limits accretion once luminosity nears a fraction of $L_{\rm Edd}$, with excess energy driving winds. 
With extra mixing (upper branch), the star remains ``immortal'' on the main sequence; without it (lower branch), the star becomes metamorphic and returns metal-rich winds to the disk.
    }
  \label{fig:schematic}
\end{figure*}

In this section, 
we develop the theoretical framework for a star evolving in an AGN–disk, 
highlighting the key physical processes: {the dependency of Helium fraction, accretion from the disk, radiative feedback, and stellar–wind mass loss}, and aim to derive simple conclusions that we can test with numerical simulations.
We begin by describing the AGN–disk environment and the baseline Bondi accretion rate onto the star (§2.1).
We then introduce how radiative feedback imposes an Eddington-limited accretion rate (§2.2) and how we implement a suppression factor $S_{\lambda_0}(\lambda_\star)$ to smoothly regulate accretion as the star approaches this limit. 
Next, 
we discuss the onset of continuum-driven stellar winds when the star’s luminosity is near Eddington (§2.3) 
and explore the condition for quasi accretion–wind equilibrium (§2.4). 
If an equilibrium is reached, the star can maintain a steady mass (potentially becoming an ``immortal" main–sequence star), 
but if not, the star will eventually enter a mass-losing post-main–sequence evolution and die 
(``metamorphic" evolution),
with important consequences for mixing and chemical yields, {as illustrated by Figure \ref{fig:schematic}}.

\subsection{Accretion in an AGN disk: Baseline Rates}
\label{sec:baselinerates}

The outer 
self-gravitating, 
star-forming region of an AGN may be approximated by a simple constant $\alpha$ and constant $Q$ disk model \citep{SirkoGoodman2003,MacLeodLin2020,ChenLin2024}.
With an $\alpha$ prescription for accretion rate,
an accretion disk attains a steady state with 
a constant accretion rate

\begin{equation}
    {\dot M}_{\rm d}= \frac{3 \alpha h^3}{Q} M_\bullet \Omega = \frac{3 \alpha c_{\rm s}^3}{G Q}
\end{equation}
where $c_{\rm s}$ is the mid-plane sound speed of the disk gas and radiation, $M_\bullet$ and 
$m_8 = M_\bullet/10^8 M_\odot$ are the mass and normalized mass of the central {super-massive black hole (SMBH)}.
In terms of the gravitational stability parameter $ Q=c_{\rm s} \Omega/ \pi G \Sigma \sim 1$, the
midplane density

\begin{equation}
    \rho_c = \frac{ M_\bullet }{ 2 \pi Q R^3}   = \frac{10^{-15}}{Q} 
    \frac{m_8}{r_{\rm pc}^3} \frac{\rm g}{cm^3}
= \frac{10}{Q m_8^2} \frac{R_\bullet ^3}{R^3} \frac{\rm g}{cm^3}
\label{eq:rhoc}
\end{equation}
where $R$ and $r_{\rm pc} = R/{\rm 1 pc}$ are the physical and normalized radius, 
$R_\bullet = G M_\bullet/c^2$ is the SMBH's gravitational radius. 
For luminous AGNs, 

\begin{equation} 
{\dot M}_\bullet \simeq 
2 m_8 \frac{\lambda_{\bullet 6}}{\epsilon_{\bullet 06}} \frac{M_\odot}{{\rm yr}}
\end{equation}
where $\lambda_\bullet = L_\bullet/L_{\rm Edd, \bullet}$ and
$\epsilon_\bullet$ are SMBH's Eddington and efficiency factors,  
$\lambda_{\bullet 6} = \lambda_\bullet/0.6$ and $\epsilon_{\bullet 06}
=\epsilon_{\bullet}/0.06$ are normalized by their mean
values inferred from the evolution of AGN's luminosity function
\citep{yu2002, shankar2009}.
SMBH's Eddington luminosity limit is $L_{\rm Edd, \bullet} 
= 4 \pi G M_\bullet  c / \kappa_e = 1.26 \times 10^{46} m_8$erg s$^{-1}$ where $\kappa_e$ is the electron scattering opacity.
If $Q$ and $\alpha$ are constant, 
the radiation-pressure dominated sound 
speed ($c_{\rm s} \approx c_{\rm s, rad}$) is nearly constant,
\begin{equation}
    c_{\rm s, rad}= \left({Q G {\dot M}_\bullet \over 3 \alpha} \right)^{1/3} =
    14  \left( {m_8 \lambda_{\bullet 6} \over \epsilon_{\bullet 06}} \right)^{1/3} 
    {\rm km \over s}.
    \label{eq:csrad}
\end{equation}
with mid-plane temperature 
\begin{equation}
    \rho_{\rm c} c_{\rm s, rad}^2 = \frac{a}{3} T_c^4.
    \label{eq:tmidplane}
\end{equation}

To verify self-consistency, we find the ratio of radiation to gas pressure in the mid-plane:
\begin{equation}
    \Pi = \dfrac{c_{\rm s, rad}^2}{c_{\rm s, gas}^2} = 
    \dfrac{\mu a ({3\rho c_{\rm s, rad}^2}/{a})^{3/4}}{3\rho \mathcal{R}} 
    =  15 Q^{1/4} m_8^{1/4} {\lambda_{\bullet 6}^{1/2} \over 
    \epsilon_{\bullet 06}^{1/2} } r_{\rm pc}^{3/4}
\label{eq:csradgasratio}
\end{equation}
increases with $R$ and exceeds unity (radiation pressure dominant) at 
$R >0.1$pc around $\sim 10^8 M_\odot$ SMBHs.

Before we consider feedback from the stellar radiative luminosity, which can be a complicated 
function of stellar mass $M_\star$ and/or radius $R_\star$, 
the stars' nominal Bondi accretion rate is given by 
 \begin{equation}
    {\dot M}_{\rm Bondi} =4 \pi R_B^2 \rho_c c_{\rm s, gas}= 4 \pi \rho_c {G^2 M_\star ^2 \over c_{\rm s, gas}^3}  
    \label{eq:mdotbondi}
\end{equation}
where $\Omega=\sqrt {G M_\bullet/R^3}=2.22 \times 10^{-10} m_8^{1/2} r_{\rm pc}^{3/2} s^{-1}$
and $R_B= GM_\star/c_{\rm s,gas}^2$ is the conventional Bondi radius. 
A point that we would like to clarify is that the sound speed governing Bondi radius, 
or the critical radius of infall (even without feedback), 
should be the gas sound speed rather than the radiation sound speed as assumed by \citet{Cantiello2021, Dittmann2021, Fabj2025}. 
This is because the optical depth $\tau$ in these regions does not exceed $c/c_{\rm s}$ and radiative diffusion timescale is still relatively short, 
therefore radiation force can act as offset to gravity to provide feedback 
\citep{Chen2024, Chen2025}. {We have neglected the stars’ relative motion with respect to the local Keplerian orbit, 
which may be comparable to the local sound speed 
and complicate the accretion structure \citep{ChenLin2024}. Nevertheless, the interplay between radiative feedback and orbital eccentricity/inclination remains to be explored with dedicated simulations.}
From Eqs. (\ref{eq:csrad}), (\ref{eq:csradgasratio}), and
(\ref{eq:mdotbondi}), we find 
\begin{equation}
    {\dot M}_{\rm Bondi} = 5 \times 10^{21} {m_8^{3/8} m_\star^2 \epsilon_{\bullet 06} ^{1/4}
    \over \lambda_{\bullet 6}^{1/4} r_{\rm pc} ^{33/8}} {\rm g \over s}
    \label{eq:mdotbondi2}
    \end{equation}
    where $m_\star=M_\star/M_\odot$.
In the alternative adiabatic regime where the optical depth $\tau\gg c/c_{\rm s}$, the Bondi radius is defined by the coupled pressure of radiation and gas, and there will not be radiative feedback altogether \citep{Chen2024}.

In the regime $c/c_{\rm s} >\tau \gg 1$, the diffusive radiation luminosity at the critical radius of infall is responsible for determining the strength of feedback. 
Simulations show that once an accretion flow with ${\dot M}_{\rm acc}$ is established, 
in addition to the stars' intrinsic luminosity $L_\star$, 
both thermal and kinetic energy are released as extra radiation with the
accreting mass ${\dot M}_{\rm acc}$ \citep{Chen2024}

\begin{equation}
    L_{\rm acc, th}({\dot M}_{\rm acc}) = {\dot M}_{\rm acc} c_{\rm s, rad}^2 = {\dot M}_{\rm acc} {a T_c ^4 \over 3 \rho_c} 
\end{equation}
\begin{equation}
L_{\rm acc, KE}({\dot M}_{\rm acc}) = {\dot M}_{\rm acc} V_\star^2 \simeq {\dot M}_{\rm acc} {G M_\star \over R_\star}
    \label{eq:lacc}
\end{equation}
where $V_\star\simeq \sqrt{G M_\star/R_\star}$ is approximately the Keplerian speed at the stellar 
surface. 
Since $V_\star \gg c_{\rm s, rad}$, 
the advection of thermal energy of the disk gas is generally negligible compared to the advection of gravitational potential energy. This is analogous to the Eddington luminosity for black hole accretion.

With an uninterrupted accretion rate 
$ {\dot M}_{\rm acc}\approx {\dot M}_B$, we can estimate that 

\begin{equation}
    L_{\rm acc, th} = 0.2 m_8^{25/24} { \alpha Q^{5/8} \over  (\epsilon/\lambda)^{5/12}}
    {m_\star^2 L_\odot \over r_{\rm pc}^{15/8}} 
\end{equation}

\begin{equation}
    L_{\rm acc, KE} = 200  (\epsilon/\lambda)^{1/4} {\alpha Q^{5/8} \over r_{\rm pc}^{15/8}} 
    {m_\star^3 \over r_\star} L_\odot 
\end{equation}
where $r_\star=R_\star/R_\odot \simeq m_\star ^{0.6}$.
Comparing the dominant gravitational term with the stellar Eddington luminosity:
\begin{equation}
    L_{\rm Edd, \star} = 4 \pi G M_\star  c/\kappa_e= 3.2 \times 10^4  m_\star L_\odot, 
\label{eq:leddington}
\end{equation}

 we find $L_{\rm acc, KE} > L_{\rm Edd, \star}$ when $M_\star \gtrsim 10 M_\odot$, 
which is when the 
accretion luminosity can provide strong feedback to reduce the accretion rate $\dot{M}_{\rm acc}$ to constrain 
$L_{\rm acc, KE}(\dot{M}_{\rm acc}) \lesssim L_{\rm Edd, \star}$.

\subsection{Detailed modeling of radiative feedback}
\label{sec:detailedrad}

When the star itself also contributes an intrinsic luminosity $L_\star$, 
the situation is more complex than the simple estimates given above. 
In the Bondi regime, $L_\star$ alone will be able to
effectively reduce the gravity felt by the ambient gas by a factor 
\begin{equation} 
\lambda_\star \equiv L_\star/L_{\rm Edd, \star} = 3 \times 10^{-5} / \Upsilon,
\label{eq:lambdastar}
\end{equation}
where the mass to light ratio $\Upsilon$ (normalized to its solar value) 
is a decreasing function of $m_\star$.
The critical or effective Bondi inflow rate becomes $(1-\lambda_\star)^2\dot{M}_B$. 
As $\lambda_\star \rightarrow 1$ with $M_\star \rightarrow 100M_\odot$,
there is a smooth transition toward the Eddington dominated regime such that
the accretion rate is capped by the energy limit 
$L_{\rm acc, KE} + L_{\rm acc, th} \lesssim L_{\rm Edd,  \star} - L_\star = (1-\lambda_\star) L_{\rm Edd,  \star}$ \citep{Chen2024}. 
Nevertheless, ${\dot M}_{\rm acc}$ should not be completely quenched so that 
$M_\star$ and $L_\star$ continue to increase. 


Formally, \citet{Chen2024} proposes a prescription for ${\dot M}_{\rm acc}$ across all regimes 
that can be approximated as:

\begin{equation}
\dot{M}_{\rm acc, formal}(\lambda_\star) 
\approx \min \left\{
\begin{aligned}
& (1 - \lambda_\star)^2 \, \dot{M}_{\rm Bondi} \\
& (1 - \lambda_\star) \, \dot{M}_{\rm Edd, KE} \\
& (1 - \lambda_\star) \, \dot{M}_{\rm Edd, th}
\end{aligned}
\right.
\label{eq:acclimitexact}
\end{equation}
with
\begin{equation}
\begin{aligned}
     {\dot M}_{\rm Edd, th} &= {L_{\rm Edd, \star} \over c_s^2}
    ={L_{\rm Edd, \star} 3 \rho_c \over a T_c^4} \\
    &= 6.3 \times 10^{27} 
    \left( {\epsilon_{\bullet 06} \over m_8 \lambda_{\bullet 6} } \right)^{1/3} m_\star {\rm g \over s}
\end{aligned}
\end{equation}

, and

\begin{equation}
    { {\dot M}_{\rm Edd, KE} = {L_{\rm Edd, \star} \over V_\star^2}
    = {L_{\rm Edd, \star} R_\star \over G M_\star}} = 6 \times 10^{22} r_\star {\rm g \over s},
\label{eq:mdotke1}
\end{equation}

When the stellar luminosity contributes significantly to the radiative feedback
process ($\lambda_\star \sim 1$), the accretion rate
${\dot M}_{\rm acc, formal}$ (Eq. \ref{eq:acclimitexact}) is mostly 
limited by ${\dot M}_{\rm Edd, KE}$
(Eq. \ref{eq:mdotke1}) with a characteristic timescale
\begin{equation}
    \tau_{\rm Edd, KE} ={M_\star \over {\dot M}_{\rm Edd, KE}} 
    = {\tau_{\rm Sal} R_{\bullet, \star} \over R_\star} 
    \simeq 10^3 m_\star ^{0.4}
    {\rm yr}
\label{eq:taueddke}
\end{equation}
where $\tau_{\rm Sal} = M_\star c^2/L_{\rm Edd, \star}=4.5 \times 10^8$yr is the Salpeter timescale and $R_{\bullet \star}$ is the
star's gravitational radius. For a solar-type star, $\tau_{\rm Edd, KE} \simeq 10^3$ yr
and it is about an order of magnitude longer for stars with $M_\star\gtrsim10^2 M_\odot$.

In practice, the stiffness of $\dot{M}_{\rm acc, formal}$ in Eq. (\ref{eq:acclimitexact})
introduces numerical instabilities.  We introduce a logistic tapering function:
\begin{equation}
    \begin{aligned}
        S_{\lambda_0}(\lambda_\star)&=\left[\dfrac12\left(1 - \tanh (4\ln\dfrac{\lambda_\star}{\lambda_0})\right)\right]^\beta\\
        &= \left[1 - {(\lambda_\star /\lambda_0)^8 \over 1 + (\lambda_\star/\lambda_0)^8}  
        \right]^\beta 
    \end{aligned}
\label{eq:qlambda}
\end{equation}
where $\lambda_0$ is the feedback-transition parameter and the power indices $\beta $ (default 2) is chosen to adjust the sharpness of the transition and to preserve numerical stability. 
The suppression factor $S_{\lambda_0} \rightarrow 1$ for $\lambda_\star \ll \lambda_0$ 
and $S_{\lambda_0} \rightarrow 0$ 
for $\lambda_\star \gg \lambda_0$. 
The radiative feedback is taken into account with
\begin{equation}
    {\dot M}_{\rm acc} (\lambda_\star)=   \min 
    \begin{cases}
        {(1-\lambda_\star)^2}{\dot M}_{\rm Bondi}\\
        S_{\lambda_0} (\lambda_\star){\dot M}_{\rm Edd, KE}\\
        S_{\lambda_0}  (\lambda_\star){\dot M}_{\rm Edd, th}
    \end{cases}
    \label{eq:acclimit}
\end{equation}

\subsection{Stellar wind mass loss}

{Luminous, hot stars} usually produce momentum dominated, line-driven winds 
\citep{lamers1999}. As $L_{\rm total} \rightarrow L_{\rm Edd, \star}$, 
opacity due to electron scattering provides a
much more effective coupling
between the radiation and matter than line opacity and can drive {continuum-driven} winds \citep{owocki2012}.

Under the assumption that a fraction of 
\begin{equation}
L_{\rm total} = L_\star + L_{\rm acc, th}+ L_{\rm acc, KE}
\label{eq:ltotal}
\end{equation}
is carried by the energy-dominated wind, we can prescribe:
\begin{equation}
\begin{aligned}
    {\dot M}_{\rm wind} &\approx (1-S_{\lambda_0}) {(L_{\star} +   L_{\rm acc, KE}  +   L_{\rm acc, th}  )\over V_{\rm escape}^2} \\
    &= (1-S_{\lambda_0}) { (L_{\star} +   L_{\rm acc, KE}  +   L_{\rm acc, th}) R_\star \over 2 G M_\star} 
    \end{aligned}
\label{eq:mdotwind}
\end{equation}
where we used $1-S_{\lambda_0}$ as a coefficient that transitions from 0 toward 1 smoothly
and $V_{\rm escape}^2 = 2 G M_\star/R_\star$. 
The mass loss intensifies and
accretion tapers down as $\lambda_\star \rightarrow 1$.

Note that it is possible for merging stars to significantly increase their $L_\star$ 
such that $\lambda_\star \sim 1$ within a few dynamical timescales. 
In this limit, $S_{\lambda_0} \simeq 0$ and the wind is launched with full intensity ${\dot M}_{\rm wind}$ (Eq. \ref{eq:mdotwind}) on a time scale

\begin{equation}
    \tau_{\rm wind}= M_\star/{\dot M}_{\rm wind} \simeq    \tau_{\rm Sal} R_{\bullet, \star}/2 \lambda_\star R_\star =     \tau_{\rm Edd, KE}/2 \lambda_\star
\label{eq:tauwind}
\end{equation}
comparable to the disk's dynamical timescale 
$\tau_{\rm dyn} = \Omega^{-1} = 1.5 \times 10^3m_8 ^{-1/2} r_{\rm pc}^{3/2}$ yr.

\subsection{Quasi accretion–wind equilibrium}
\label{sec:accretwind}

Taking both accretion and wind into account, the stellar mass evolves 
with a net rate
\begin{equation}
    {\dot M}_{\rm net} = {\dot M}_{\rm acc} - {\dot M}_{\rm wind}.
\label{eq:mdotnet}
\end{equation}
When a quasi accretion–wind equilibrium in which $\dot{M}_{\rm acc} 
= \dot{M}_{\rm wind}$ is established with 
$\lambda_\star = \lambda_{\rm equi}$, $S_\lambda 
{\dot M}_{\rm Edd, KE}$ usually sets the 
limit on ${\dot M}_{\rm acc}$ in Eq.~(\ref{eq:acclimit}). 
Combining with  Eq.~(\ref{eq:lacc}) Eq.~(\ref{eq:mdotke1}) and 
(\ref{eq:mdotwind}), we find an equilibrium at
$\lambda_\star \simeq \lambda_{\rm equi} \simeq \lambda_0$ due to the tapering form of $S_{\lambda_0}$. 
The evolution tracks for immortal stars of any initial mass are expected to converge to this state, 
before metamorphic stars eventually deviate 
due to exhaustion of hydrogen.

{This net effect is, of course, only a crude approximation to a potential quasi-steady state. In the isotropy limit, the wind stalling radius $R_{s}$ for a given $\dot{M}_{\rm wind}$ is given by}

\begin{equation}
   \dfrac{\dot{M}_{\rm wind} v_\infty}{4\pi R_{s}^2} = \rho c_s^2,
\end{equation}

{assuming the wind terminal velocity $v_\infty$. We also note that $\dot{M}_{\rm acc} = 4\pi \rho c_s R_{\rm crit}^2$, where $R_{\rm crit}$ is the actual critical radius of infall where inward velocity crosses local sound speed, determined by the radiative feedback strength \citep{Chen2024}. Then}

\begin{equation}
    \dfrac{R_{s}}{R_{\rm crit}} =  \left({\dfrac{\dot{M}_{\rm wind} v_\infty}{\dot{M}_{\rm acc}c_s}}\right)^{1/2}
\end{equation}

{for $v_{\infty}$ close to the escape velocity of a massive star $\sim 200$km/s, $\sqrt{v_{\infty}/c_s} \sim 2-3$. 
this demonstrates that when the mass loss rate is significantly lower than accretion $\dot{M}_{\rm wind} \ll \dot{M}_{\rm acc}$, the wind is quenched. Conversely, in the opposite limit $\dot{M}_{\rm wind} \gtrsim \dot{M}_{\rm acc} \sim \dot{M}_{\rm Edd} $ the wind bubble expands beyond $R_{\rm crit}$ and pushes material outward, thereby suppressing accretion \citep{Liu2025,Liu2025b}. 
Our taper function is introduced partly to capture this transition, where wind and accretion may coexist in the form of porous stellar winds. Additionally, there may be anisotropies that accommodate different characteristic solid angles for accretion and wind \citep{Chen2025}. }

\begin{deluxetable*}{lcc|cccc}
\tablecaption{Model parameters and summary data. We show metamorphic stars with standard mixing prescription and immortal stars with extra mixing prescription (\S \ref{sec:mixing}). 
}
\label{tab:grid_mixing_all}
\tablehead{
\colhead{$\lambda_0$} & \colhead{$Y_{\rm d}$[\%]} & \colhead{$\rho_{\rm c}$[cgs]} &
\colhead{$M_{\max}[M_\odot]$} & \colhead{$M_{\rm Hdep}[M_\odot]$} & \colhead{$\log \dfrac{\tilde L_{1+2}}{L_\odot}$
\tablenotemark{a}} & \colhead{$\log \dfrac{\tilde L_{2}}{L_\odot}$\tablenotemark{b}}
}
\startdata
\multicolumn{7}{c}{Standard Mixing}\\
\hline
{0.25} & 0.25 & $10^{-16}$   &  58.80 &  6.81 & 5.074 & 5.292\\
{0.50} & 0.25 & $10^{-16}$   & 122.1 & 10.46 & 5.276 &5.810\\
{0.75} & 0.25 & $10^{-16}$   & 234.3 & 14.75 & 5.649 &6.113\\
{0.90} & 0.25 & $10^{-16}$   & 337.9 & 17.58 & 5.730 &6.262\\
0.50 & 0.25 & $\mathbf{10^{-17}}$  & 82.48 &  9.83 & 4.725 &5.529  \\
0.50 & 0.25 & $\mathbf{10^{-16}}$   & 122.1 & 10.46 & 5.276 & 5.810\\
0.50 & 0.25 & $\mathbf{10^{-15}}$  & 132.9 & 12.33 & 5.449 & 5.772\\
0.50 & 0.25 & $\mathbf{10^{-14}}$  & 128.8 & 13.66 & 5.505 & 5.771 \\
0.50 & 0.25 & $\mathbf{10^{-13}}$  & 127.1 & 14.05 & 5.510 & 5.774\\
0.75 & {0.30} & $10^{-16}$  & 186.4 & 14.20 & 5.619 &6.020\\
0.75 & {0.40} & $10^{-16}$  & 132.8 & 14.53 & 5.608 &5.938\\
0.75 & {0.50} & $10^{-16}$  &  94.40 & 14.46 & 5.557 &5.866\\
0.75 & {0.60} & $10^{-16}$  &  63.97 & 14.65 & 5.457 &5.762\\
0.75 & {0.70} & $10^{-16}$  &  40.02 & 14.80 & 5.317 &5.632\\
\hline
\multicolumn{7}{c}{Extra Mixing}\\
\hline
{0.50} & 0.25 & $10^{-16}$ & 122.18 & \nodata &6.246 &\nodata\\
{0.75} & 0.25 & $10^{-16}$ & 234.28 & \nodata &6.666 &\nodata\\
{0.90} & 0.25 & $10^{-16}$ & 337.98 & \nodata &6.875 &\nodata\\
\bottomrule
\enddata
{\footnotesize
\tablecomments{a}{$\log \tilde L_{1+2}$: average luminosity from model start to central-H depletion (or to the last model if H–depletion is not reached).}
\tablecomments{b}{$\log \tilde L_{2}$: average luminosity from peak mass to TAMS/the last model.}}
\end{deluxetable*}

\section{Methodology}
\label{sec:methods}

In this section, we outline some details of the simulation setup, 
including modifications for AGN–disk accretion, wind-loss, chemical mixing, 
as well as our explored parameter space.

\subsection{Code Setup}\label{sec:code}
{ We study stellar evolution in AGN disks with continuous accretion and mass loss using \texttt{MESA} v22.11.1 \citep{MESA2011,MESA2013,MESA2015,MESA2019,MESA2023Jermyn}, extending the AGN star models of \citet{Cantiello2021,Jermyn2021}.
We use the nuclear network \texttt{approx21} to follow evolution up to silicon burning. 
\texttt{MESA} adopts OPAL opacities \citep{Iglesias1996}, supplemented by \citet{Ferguson2005,Poutanen2017}, with conduction opacities \citep{Cassisi2007,Blouin2020}. 
Reaction rates combine JINA, REACLIB, NACRE and weak rates \citep{Cyburt2010,Angulo1999,Fuller1985,Oda1994}, include screening \citep{Chugunov2007}, and neutrino losses \citep{Itoh1996}.
Time resolution $<10^{10}\,\mathrm{s}$ is imposed for stability during rapid mass change.}

\subsection{Model Parameters and Boundary Conditions}
\label{sec:parameters}
At the onset of each calculation, the zero-age main-sequence (ZAMS) star has 
$M_\star=M_\odot$ and $Z = Z_\odot$. The stellar surface pressure and temperature 
are set to be those of the local disk mid-plane. Our model parameters 
(Table \ref{tab:grid_mixing_all}) span $\rho_{\rm c} =10^{-17}$–$10^{-13}$\,
g\,cm$^{-3}$, $c_{\rm s, gas}\approx 10^6$~cm~s$^{-1}$, 
and $T_{\rm c} \simeq 10^5$~K (Eq. \ref{eq:tmidplane}). 
These values are appropriate for various radii around SMBH with 
different masses (\S\ref{sec:baselinerates}).

For the abundances of the accreted disk gas, we explore $Y_{\rm d}=0.25–0.7$,
$Z_{\rm d, C}= 2.2 \times 10^{-3}$, $Z_{\rm d, N}= 7 \times 10^{-4}$, 
and $Z_{\rm d, O}= 6.3 \times 10^{-3}$ for C, N, and O respectively.

\subsection{Accretion and Wind-loss Implementation}
\label{sec:accimplementation}

At every timestep, we separately compute the accretion and continuum-driven wind rates, 
taking into account radiative-feedback with $\beta=2$ \footnote{$\beta = 1, 3$ and the results are not sensitive to this hyperparameter.}, $\lambda_0=0.25$–0.90
in the tapering function $S_{\lambda_0}$ (Eq. \ref{eq:qlambda}). 
With $S_{\lambda_0} \rightarrow 0$ as $\lambda_{\star} \rightarrow 1$, 
super-Eddington inflow is quenched. 

Accretion and wind occur concurrently with the same $S_{\lambda_0}$.
Wind loss is implemented first with ${\dot M}_{\rm wind}$ from 
Eq. (\ref{eq:mdotwind}).  
For $|\dot{M}_{\text {wind}}| 
\Delta t$ smaller than the outermost cell mass, gas in the 
surface cell is removed entirely; the excess fraction is then the replenishment
of the excess fraction with a fixed composition of the circumstellar
gas. If the amount to be removed exceeds the surface cell, successive 
outer cells are stripped until the total removed mass matches 
$|\dot{M}_{\text {wind}}| \Delta t$. 

Treatment of the wind loss is followed by applying accretion (${\dot M}_{\rm acc}$) to the exposed layers using the composition of 
the circumstellar gas. This sequence leads to ${\dot M}_{\rm net}$ in
accordance with Eq. (\ref{eq:mdotnet}).
Typical net mass loss rates stay within $10^{-12}-10^{-4} M_{\odot} \mathrm{yr}^{-1}$, 
enforced by a timestep cap of $\Delta t \leq 10^{10} \mathrm{~s}$ to preserve numerical 
stability throughout the post-main–sequence phase.

The star–disk mass exchange may modify the abundance of the circumstellar
gas. In most models, we assume total decoupling 
between the accretion flow and the outgoing wind, meaning that accreted gas carries the same composition as the disk initial condition.  We also briefly consider the 
possibility of total retention \citep{AlidibLin2023}, i.e., the stars accrete 
{\it in situ} the gas polluted by their own wind in \S \ref{sec:retention1}.

\subsection{Chemical Mixing Diffusivity}
\label{sec:mixing}

{Along the main sequence evolution, stars contain both radiative cores and 
convective envelopes. We use the prescription in \citep{Xu2025a} for convective zones. 
The metamorphic stellar models 
are generated by setting a ``standard" uniform level for minimal diffusivity 
$D_{\rm mix, rad} = 10^{5-7}\,\textrm{cm}^2\,\textrm{s}^{-1}$ for the radiative envelopes, 
representative values consistent with rotational mixing \citep{Spruit2002,Maeder2010,Prat2014}.
The immortal stars are constructed with an {\it ad hoc} ``extra-mixing'' 
prescription, in which $D_{\rm mix}(r)$ is assumed to increase with 
stellar radiative flux $F$ \citep{Cantiello2021}. }

\section{Results}\label{sec:res}

\begin{figure*}[hbtp]
  \centering
  \gridline{
    \fig{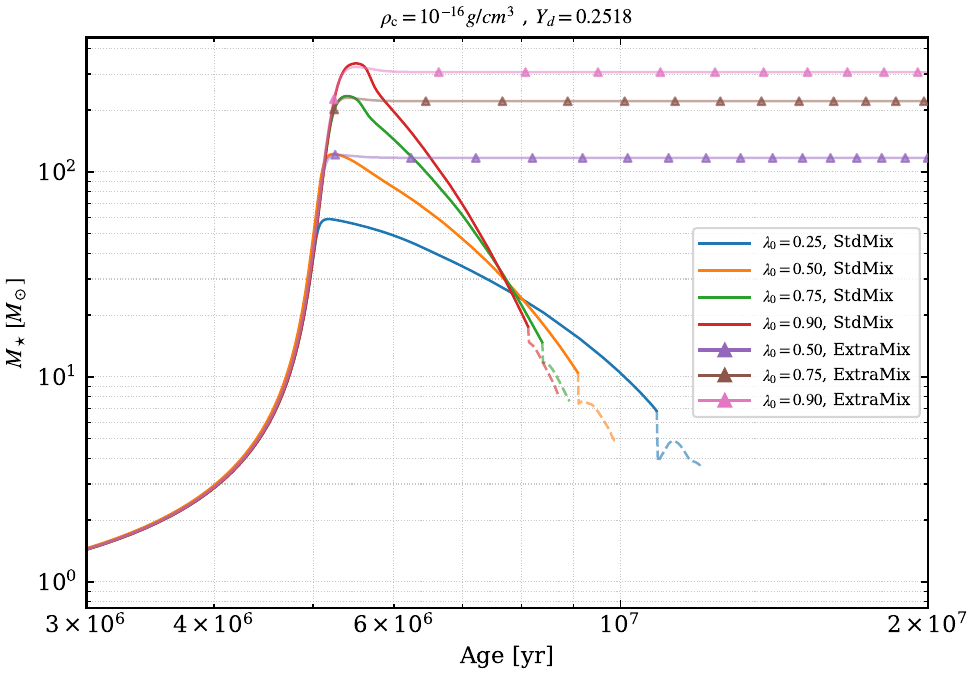}{0.45\textwidth}{(a) Mass evolution for different $\lambda_0$}
    \fig{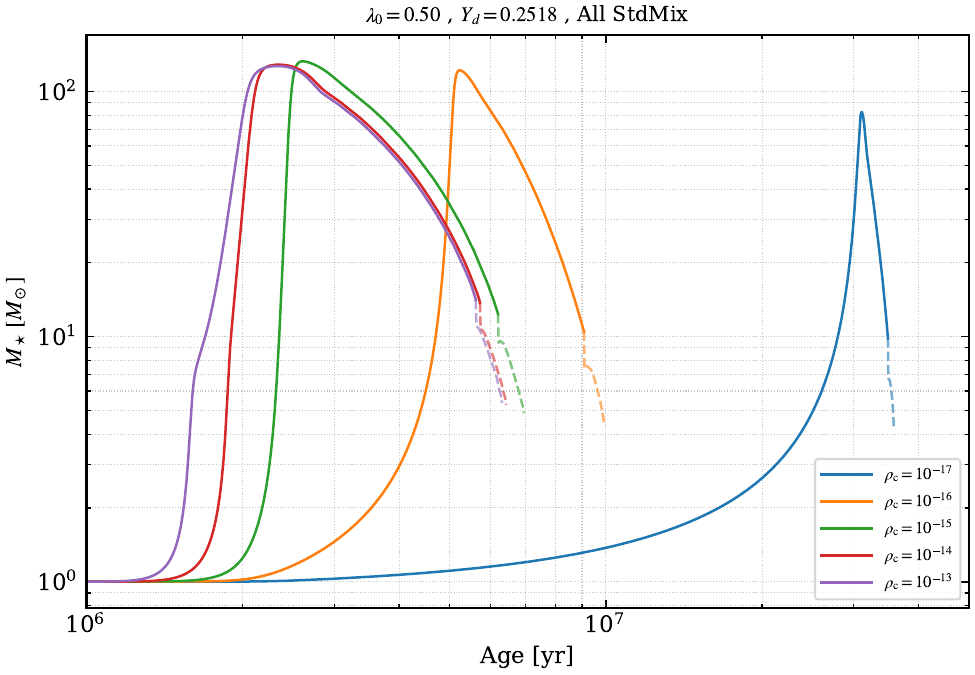}{0.45\textwidth}{(b) Mass evolution for different $\rho_{\rm c}$}
  }
  \vspace{-0.25em}
  \gridline{
    \fig{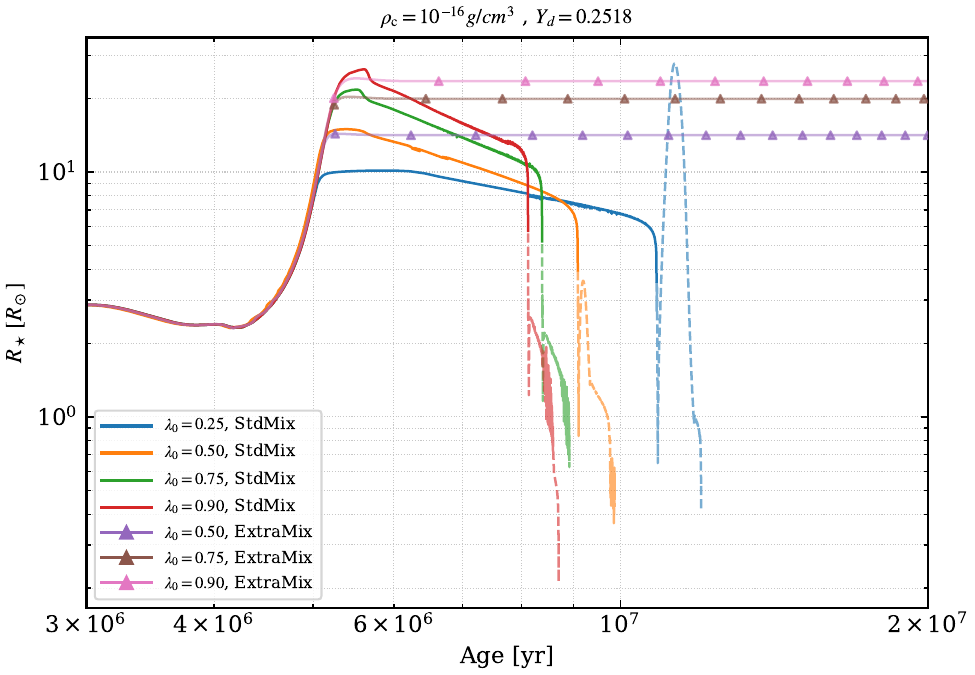}{0.45\textwidth}{(c) Radius evolution for different $\lambda_0$}
    \fig{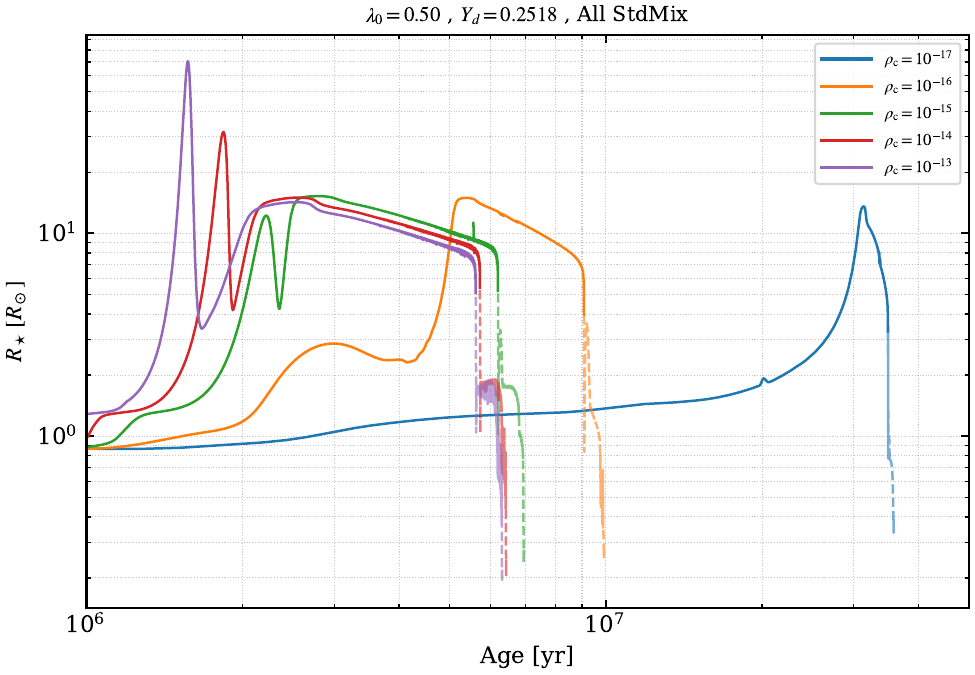}{0.45\textwidth}{(d) Radius evolution for different $\rho_{\rm c}$}
  }
\caption{
 Stellar mass (top) and radius (bottom) evolution. 
Panels (a,c) vary feedback parameter $\lambda_0$; panels (b,d) vary disk density. 
Solid lines show standard mixing; triangles denote extra mixing (immortal track). 
Feedback limits runaway growth, larger $\lambda_0$ allows higher masses, while ambient density has little effect.
} 
\label{fig:mass-growth}
\end{figure*}

\subsection{Initial Growth: Dependence on $\lambda_0$ and $\rho_{\rm c}$}

Below we present the outcomes of our \texttt{MESA} calculations
over a range of model parameters (Table \ref{tab:grid_mixing_all}),
and derive  various quantities from these simulations.

Figure \ref{fig:mass-growth} 
shows the evolution of stellar mass as a function of time, 
for several sets of representative models. 
During the initial evolution, $L_{\rm total} < L_{\rm \star, Edd}$
(Eqs. \ref{eq:leddington} \& \ref{eq:ltotal})
such that $S_{\lambda_0} \sim 1$, ${\dot M}_{\rm acc} \gg {\dot M}_{\rm wind}$
(Eqs. \ref{eq:qlambda}, \ref{eq:acclimit}, \& \ref{eq:mdotwind}),
and the stellar mass $M_\star$ undergo runaway growth, 
with a rate ${\dot M}_{\rm acc} \sim {\dot M}_{\rm Edd, KE}$ 
(Eq. \ref{eq:acclimit}) which is {\it independent} of $\rho_{\rm c}$ 
(Eq. \ref{eq:mdotke1}). Panel b in Fig. \ref{fig:mass-growth} shows that 
$M_\star$ ($\gtrsim 10 M_\odot)$ grows at similar rates across most 
($R \lesssim$ a few pc) regions of the disk.

The radiation feedback becomes strong enough to suppress accretion
to intensify mass loss through stellar wind with $\lambda_\star \rightarrow 
\lambda_0$ and $S_{\lambda_0} < 1$ as $M_\star$ reaches above a few tens $M_\odot$.
This transition marks the onset of an accretion–wind equilibrium (with 
${\dot M}_{\rm acc} \simeq {\dot M}_{\rm wind}$), which limits 
the star's initial growth rate with a maximum $M_\star = M_{\rm max}$
with $\lambda_\star \simeq \lambda_0$.

In general, $L_\star$ and $\lambda_\star$ are functions of $M_\star$ 
and $Y_\star$ \citep{owocki2012}, such that the value of $ \lambda_0$ 
can be used to infer a stellar mass $M_\star=M_{\rm equi}(\lambda_0, Y_\star)$ 
for quasi accretion–wind equilibrium as a function of the time-dependent helium 
abundance $Y_\star$, see \S~\ref{sec:evolvingmass}.  During the initial 
accretion phase, $Y_\star\sim Y_d$ so that $M_{\rm max}$ depends on both $\lambda_0$
and $Y_{\rm d}$ (panel a, Figs. \ref{fig:mass-growth} \& \ref{fig:yd-evolution}). 
After reaching $M_{\rm max}$, the evolution tracks of immortal and metamorphic stars begin to dramatically diverge. {Since with radiative feedback the accretion rate is regulated to be at most comparable to the Eddington-limited mass-loss rate, 
non of our models reach the runaway growth track in \citet{Fabj2025} where mass growth timescale becomes shorter than the stellar thermal timescale.}

\subsection{Subsequent Evolution of Immortal vs. metamorphic stars: Dependence on Mixing}

In models with extra mixing imposed in the radiative zone (triangular symbols), 
the star attains an asymptotic mass $M_{\rm IMS}$ with a slightly sub-Eddington luminosity
$L_{\rm IMS}$ when the accretion and wind rates cancel each other. 
In the full-retention limit, the circumstellar gas is locally contaminated by the
stellar wind to have the same composition as the stellar surface, 
all stars must be metamorphic due to retention of materials from the stellar wind, \textit{even} 
with flux-induced extra-mixing \citep{AlidibLin2023}.  

In this paper, we present most models with zero retention such that 
the outer mass-exchange region of the star is continually refreshed 
with the abundance of the disk gas. The outer stellar envelope contains
radiative zones which separates the nuclear burning core from the
mass exchange region. With ``extra mixing'', \citet{Cantiello2021} showed 
the core would be replenished with the H–laden accreted gas and purged 
its He ashes to indefinitely prolong these ``immortal'' stars' main–sequence evolution.  
A suite of \texttt{MESA} models in \citet{Xu2025a} shows that a radiative–zone diffusivity floor of \(D_{\rm mix,rad}\gtrsim 10^{10}\,\mathrm{cm^{2}\,s^{-1}}\) leads to ``immortality''. 

For models with the standard mixing prescription and various values of $\lambda_0$ 
and $Y_{\rm d}$, the stellar mass reaches  maximum values $M_{\rm max}$ similar to immortal stars with corresponding model parameters.  
However, with an inadequate H diffusion and replenishment to the core (even for 
our assumption of zero–retention of stellar wind materials), quasi accretion–wind 
equilibrium is maintained with an increasing $Y_\star$ and decreasing $M_\star$.
Details of this phase is elaborated in \S\ref{sec:evolvingmass}
where we discuss models with different $Y_d$ to aid our analysis. 

After exhausting H in the convective core, the star undergoes He 
burning through triple–$\alpha$ process and undergoes a transition
to post main–sequence evolution. The doubling of its luminosity 
(above $L_{\rm Edd, \star}$) leads to fractional reduction of its
mass with a modest expansion of its radius (Fig. \ref{fig:mass-growth}). 
This transition of the metamorphic
stars in AGN disk is in contrast to the red–giant phase of stand–alone
stars in the Galaxy. An exception is the $\lambda_0=0.25$ (strong–feedback)
model which produced a red–giant expansion followed by a $R_\star$ contraction
after the onset of He burning. Finally, the depletion of He leads to
the nuclear burning of other elements until the star undergo core–collapse,
leaving behind a black hole of mass perhaps $5\sim 15\, M_\odot$,
with or without type II supernova \citep{Fryer2025}.

\begin{figure}
  \centering
  \includegraphics[width=0.9\linewidth]{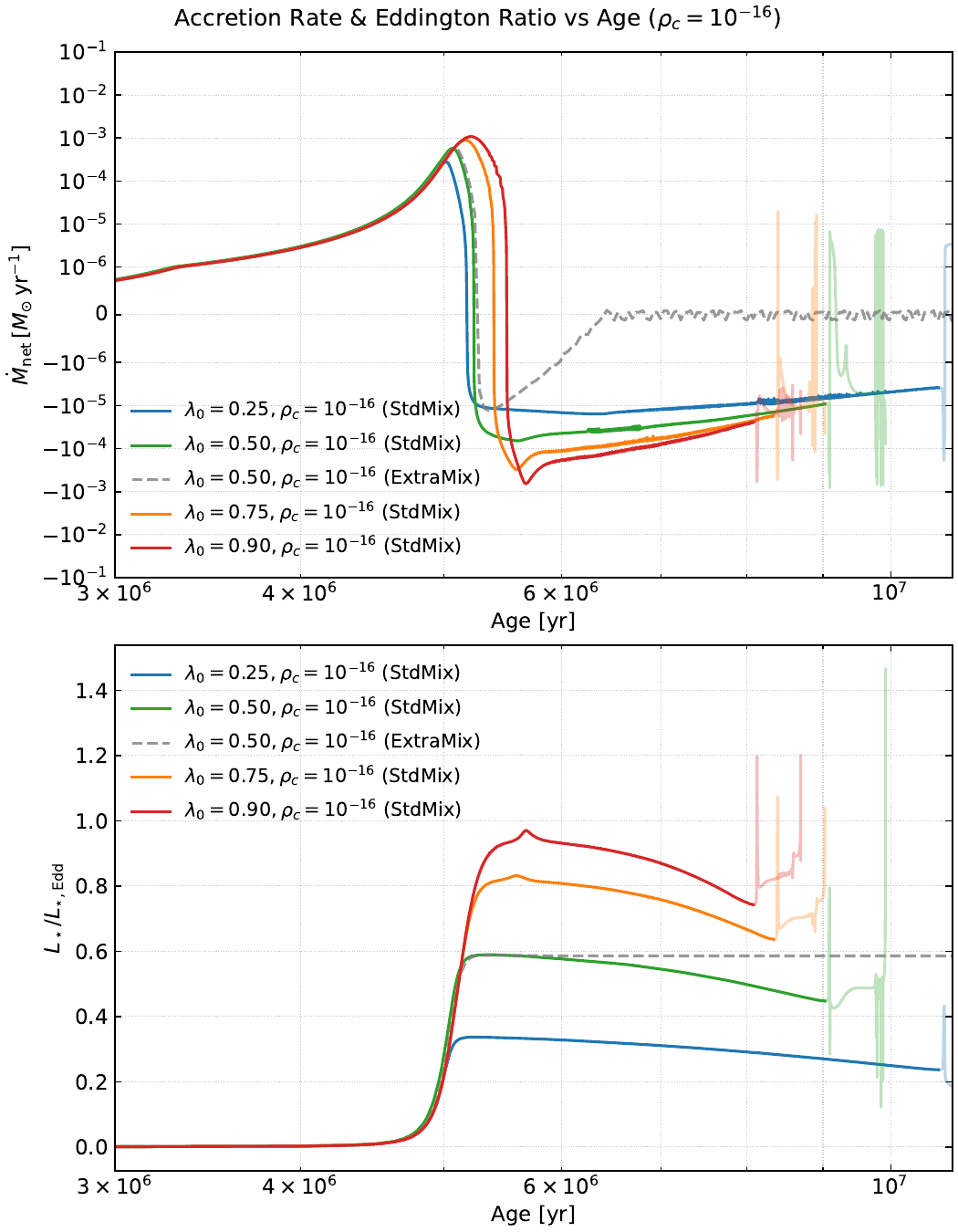}
  \caption{
    Evolution of net mass–exchange rate
    $\dot M_{\rm net}$, i.e. the accretion rate deducts the absolute value of the 
    wind–loss rate is shown in the upper panel. The 
    Eddington ratio $L_\star/L_{\mathrm{Edd, \star}}$ is shown in the lower panel
    for the fiducial
    model with $\rho_{\mathrm{c}}=10^{-16}\, \mathrm{g\,cm^{-3}}$, $Y_{\rm d} = 0.25$, and
    various values of $\lambda_{0}$.  Opaque curves indicate main–sequence
    and partially transparent curves represent post–main–sequence phases after core hydrogen exhaustion.  
    The dashed line represents the immortal–star model with $\lambda_0 = 0.5$ and extra mixing.}
  \label{fig:mdot-lledd-age}
\end{figure}


Figure \ref{fig:mdot-lledd-age} illustrates the time evolution of ${\dot M}_{\rm net}$ and 
$\lambda_\star$ for models with a fixed $\rho_{\rm c}= 10^{-16}$g/cm$^3$, highlighting the 
difference between immortal and metamorphic stars.  During the first $\sim$5 Myr, 
$\dot{M}_{\rm acc}$ ramps up to its maximum (on the order of $10^{-4} - 10^{-3} 
M_\odot$/yr).  Since $L_{\star} < \lambda_0 L_{\rm Edd, \star}$, $\dot{M}_{\rm wind}$ 
is negligible during this growth phase. The Eddington ratio $\lambda_\star $ rises as the 
$M_\star$ grows (Eq. \ref{eq:lambdastar}). At $t \approx 5$ Myr, $\lambda_\star \rightarrow
\lambda_0$ and radiative feedback begins to quench $\dot{M}_{\rm acc}$ and 
$\dot{M}_{\rm net}$ decreases precipitously after 5 Myr. For metamorphic stars, 
$\dot{M}_{\rm net}$ becomes negative during this He–enriching main–sequence phase. For the immortal star example ($\lambda_0= 0.5$) 
shown in Figure \ref{fig:mdot-lledd-age},  the net mass loss reduces to zero 
due to continuous exchange of composition with the disk background and both 
$M_\star$ and  $\lambda_\star$ are maintained at constant values.

\subsection{Evolution of equilibrium mass: Effect of $Y_d$ and $Y_{\star}$}
\label{sec:evolvingmass}

\begin{figure*}[ht]
\centering
  \gridline{
    \fig{\detokenize{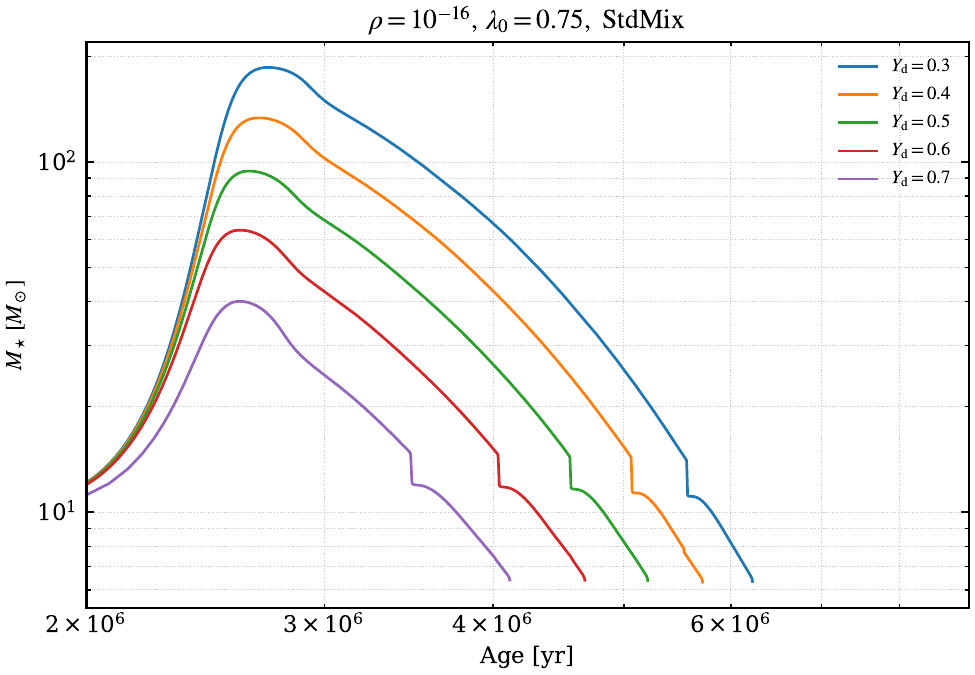}}{0.45\textwidth}{(a) Stellar mass $M_\star$ evolution for different disk helium‐fraction$Y_d$}
    \fig{\detokenize{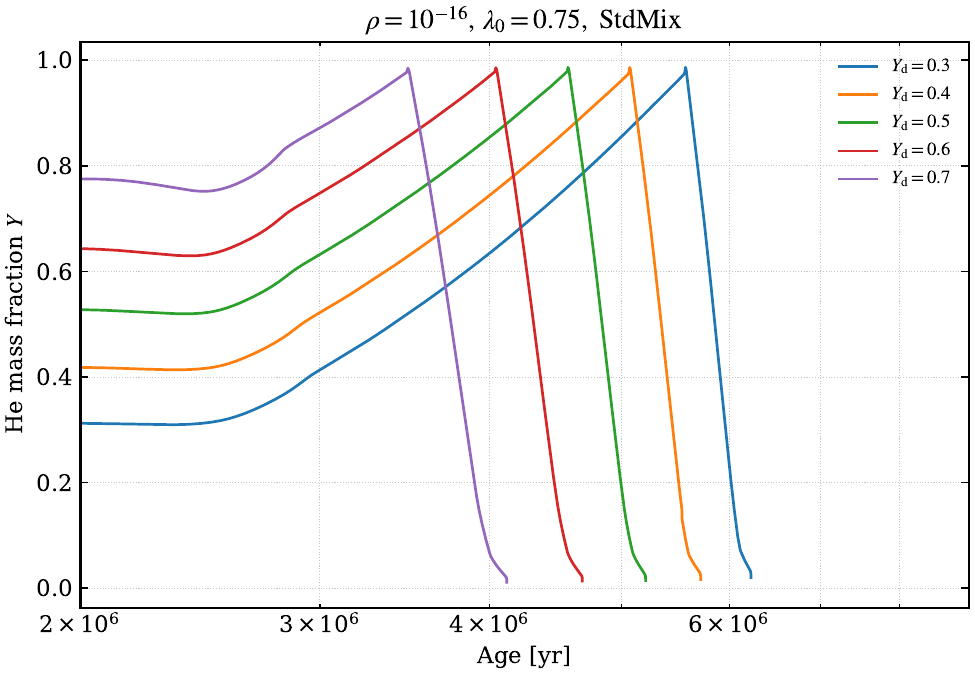}}{0.45\textwidth}{(b) Stellar helium‐fraction $Y_\star$ evolution for different disk helium‐fraction $Y_d$}
  }
  \vspace{-0.25em}
  \gridline{
    \fig{\detokenize{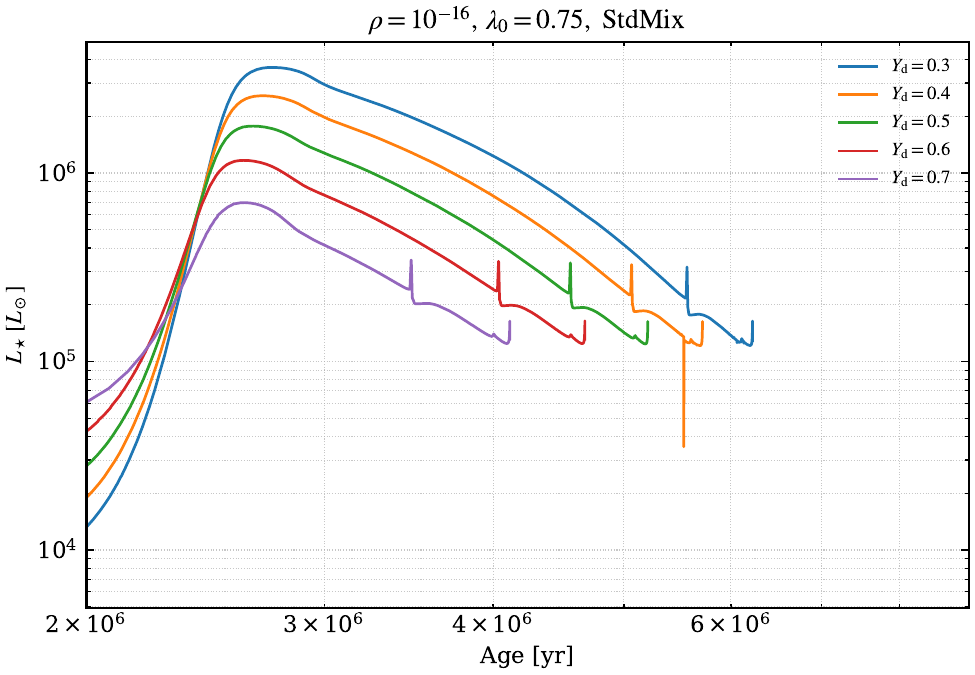}}{0.45\textwidth}{(c) Luminosity evolution for different $Y_d$}
    \fig{\detokenize{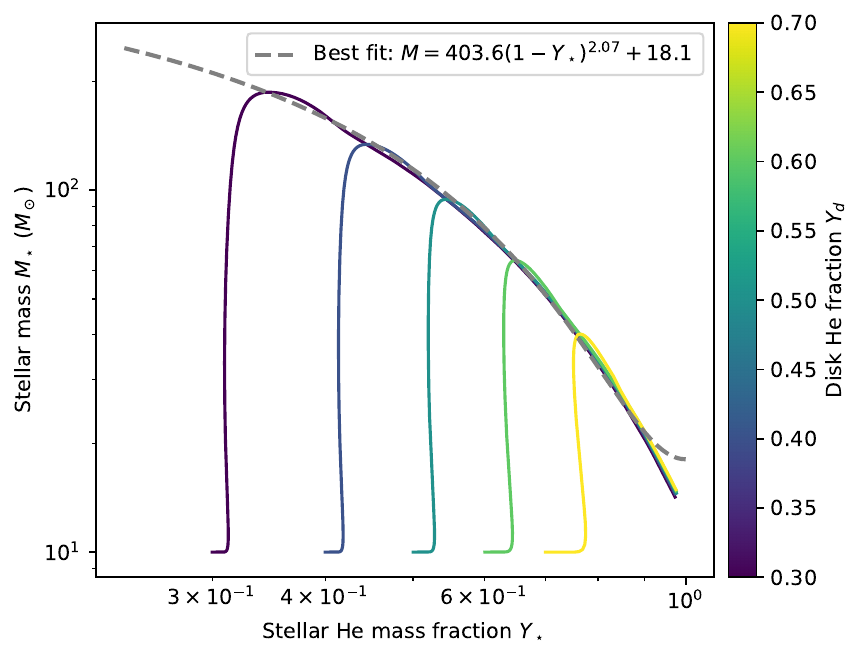}}{0.45\textwidth}{(d) Stellar $Y_\star$ vs $M_\star$ relation with power–law fit}
  }
\caption{Stellar evolutionary tracks for varying disk helium mass fractions $Y_{\rm d}=0.3$–$0.7$, with fixed $\rho_{\rm AGN} = 10^{-16} \mathrm{g\,cm^{-3}}$ and $\lambda_0 = 0.75$. Panels show the evolution of (a) stellar mass $M_\star$, (b) helium mass fraction $Y_\star$, (c) luminosity $\log L_\star$, and (d) the $Y_\star$–$M_\star$ relation during main–sequence evolution. As $Y_{\rm d}$ increases, higher mean molecular reduces the mass–to–light ratio $\Upsilon$, causing stars to reach the feedback limit  with lower final masses. Panel (d) shows the best–fit power–law relation (grey dashed) between $Y_\star$ and $M_\star$ during equilibrium growth, as predicted by Eq.~(\ref{eq:y_time}). Because models start from 10$M_\odot$, the ramp–up time is shortened.}
\label{fig:yd-evolution}
\end{figure*}

Figure \ref{fig:yd-evolution} shows the evolution track for different helium environmental abundance. 
A very clear trend is that the overall mass scales down with $Y_d$, 
offering insight into the mechanism driving quasi–steady mass loss following the initial growth phase.
Since $\tau_{\rm acc, KE} \ll \tau_{\rm He}$ (Eqs. \ref{eq:taueddke} and \ref{eq:tauhe}), 
the helium fraction in the star $Y_\star \sim Y_d$ when the stars have just 
reached their maximum mass (lower right panel).
Due to the dependence of stars (or more generally stellar population's) mass–to–light ratio $\Upsilon (M_\star, Y_\star)$
on the average molecular weight\citep{owocki2012, AlidibLin2023}, 
the maximum mass at initial equilibrium
$M_{\rm max} (\lambda_0, Y_d) \simeq M_{\rm equi}(\lambda_0, Y_\star \simeq
Y_d)$ for a given $\lambda_0$ scales down with $Y_{\rm d}$ (or equivalently, 
up with $X_{\rm d}$).  

After this point, $Y_\star$ and the average molecular weight increase with time.  
Since $\Upsilon (M_\star, Y_\star)$ is a decreasing function of
$M_\star$ and $Y_\star$, the preservation of the quasi accretion–wind equilibrium with $\lambda_\star \sim \lambda_0$ and 
\begin{equation} 
\Upsilon_{\rm equi} = 3 \times 10^{-5} / \lambda_0,
\label{eq:upsiloneq}
\end{equation}
from Eq. (\ref{eq:lambdastar}), requires $M_\star$
to decrease with increases in $Y_\star$  (Fig. 3 in \citet{AlidibLin2023})
on the hydrogen depletion timescale $\tau_{\rm He}$ (Eq. \ref{eq:tauhe}) 
in the nuclear burning core,  regardless the initial helium contents.
Moreover, $L_\star$ of massive stars increases with $M_\star$
slightly steeper than a linear relation so their $\Upsilon$ 
decreases slowly with $M_\star$ and small changes in $\lambda_0$
and $\Upsilon_{\rm equi}$ (Eq. \ref{eq:upsiloneq}) lead to notable 
modification to the equilibrium mass $M_{\rm equi}$.
 
To model the time-dependence of stellar mass in this quasi-steady mass loss phase, we may assume $\lambda_\star\sim \lambda_0$ and write the energy equation as 
\begin{equation}
    \epsilon_{\rm He} {\dot M}_{\rm H} c^2 \simeq \lambda_0 L_{\rm Edd, \star} 
\end{equation}
where $ {\dot M}_{\rm H}$ is the hydrogen consumption rate and $\epsilon_{\rm He}
\sim 0.007$ is the conversion efficiency from rest-mass energy into radiation.  
In the CNO cycle on the main sequence, He is enriched at the same rate as 
H is depleted, i.e. ${\dot M}_{\rm He} = {\dot M}_{\rm H}$ and ${\dot Y}_\star
=-{\dot X}_\star$.
The characteristic timescale for H depletion (or equivalently He enrichment) in the nuclear 
burning core (with a mass $M_{\rm core} \lesssim M_\star$) is 

\begin{equation}
    \tau_{\rm He} \simeq {X_\star M_{\rm core} \over {\dot M}_{\rm H}} 
    \simeq {X_\star \epsilon_{\rm He} \tau_{\rm Sal} \over \lambda_0} {M_{\rm core} \over M_\star}
    \sim 3 {X_\star M_{\rm core} \over \lambda_0 M_\star} {\rm Myr}
\label{eq:tauhe}
\end{equation}
where $\tau_{\rm Sal} \simeq M_\star c^2 / L_{\rm Edd, \star}$ is the Salpeter timescale.

In massive stars, most of the mass is contained in the convective core, i.e. $M_{\rm core} 
\sim M_\star$ so that the normalized enrichment rate  becomes
\begin{equation}
    {\dot Y}_\star= {{\dot M}_{\rm H} \over M_{\star}} 
    \simeq {\lambda_0 \over \epsilon_{\rm He} \tau_{\rm Sal}}  
    \label{eq:mhedot}
\end{equation}
which leads to a linear increase of $Y_\star$
(with a universal slope independent of the value of initial $Y_d$, lower right panel) with duration of time after reaching maximum mass at $t_0$: 
\begin{equation}
    { Y}_\star 
    \simeq Y_d + {\lambda_0 (t-t_0) \over \epsilon_{\rm He} \tau_{\rm Sal}},
    \label{eq:yevolve}
\end{equation}

Since $\tau_{\rm He} \gg \tau_{\rm acc, KE}$ and $\tau_{\rm He} \gg \tau_{\rm wind}$ 
(Eqs. \ref{eq:taueddke} and \ref{eq:tauwind}), a quasi-steady mass loss state is maintained.
As $Y_\star$ increases with He enrichment, $M_\star (Y_\star)$ converging to the intrinsic 
values of $M_{\rm equi}(\lambda_0, Y_\star)$. 
For a given $\lambda_0$, the results of 
the \texttt{MESA} models can be fitted (Bottom right panel in Fig.~\ref{fig:yd-evolution}) as 
\begin{equation}
    M_{\rm equi} (\lambda_0, Y_\star)\simeq M_{\star, \rm H}(\lambda_0)(1- Y_\star)^2 + M_{\star, \rm He}(\lambda_0)
    \label{eq:y_time}
\end{equation}
where $M_{\star, \rm H}(\lambda_0)$ and $M_{\star, \rm He}(\lambda_0)$ are numerical values that can be interpreted as 
the mass of a fully hydrogen 
and fully helium star at Eddington ratio $\lambda_0$. This fit satisfies both 
$\lambda_\star \simeq \lambda_0$, and the stellar mass-to-light $\Upsilon (M_\star, 
Y_\star) = \Upsilon_{\rm equi}$ (Eq. \ref{eq:upsiloneq}) relation.  
From Bottom right panel in Fig.~\ref{fig:yd-evolution}, 
we can fit $M_{\star, \rm H} \sim 400 M_\odot$ and $M_{\star, \rm He} \sim 18 M_\odot$
for $\lambda_0=0.75$, so the mass \textit{time-dependence} after $t_0$ can be described by 
\begin{equation}
    \begin{aligned}
        &M_\star = M_{\rm equi}(\lambda_0, Y_d, t -t_0) \\
        & \simeq  M_{\star, \rm H}(\lambda_0)+ M_{\star, \rm He}(\lambda_0)\left(1- Y_d - {\lambda_0 (t-t_0) \over \epsilon_{\rm He} \tau_{\rm Sal}}\right)^2
    \end{aligned}
    \label{eq:mstarlambday}
\end{equation}

When H in their core is exhausted (with $Y_\star \rightarrow 1$), stars undergo 
transition to post-main–sequence with a $M_\star$ which is greatly reduced from 
its maximum values shortly after their formation. For $\lambda_0 \sim 0.5-0.9$,
$M_\star$ ranges from $10.5 M_\odot-17.7 M_\odot$ at the stage of H exhaustion in the core.

\subsection{Stellar Population}

In contrast to stand-alone stars in the Galaxy,
metamorphic stars in AGN disks with the same age ($t-t_0$) have the same values of 
$Y_\star$ (Eq. \ref{eq:yevolve}), $M_\star (\simeq M_{\rm equi})$ (Eq. \ref{eq:y_time}), 
$\lambda_\star (\simeq \lambda_0$), and $L_\star (\simeq \lambda_0 L_{\rm Edd, \star})$.
At any given time $t$, their luminosity and mass functions are determined by their
formation epoch $t_0$. 

We partition the evolution into four phases: 
Phase 1—initial ramp to the maximum mass $M_{\max}$ with $\dot{M}_{\rm acc}\gg\dot{M}_{\rm wind}$; 
Phase 2—$M_{\max}$ to central-H depletion; 
Phase 3—central-H depletion (onset of He burning) to central-He depletion (onset of C burning); 
Phase 4—central-He depletion (C burning) to pre-collapse (Si burning).

\subsubsection{Luminosity and Energy Output}

During their finite lifespan $\tau_\star ( \sim {\rm a \ few \ Myr \ } 
\geq \tau_{\rm He}$), metamorphic stars 
have an average luminosity
\begin{equation}
    {\tilde L}_{\rm total} = {1 \over \tau_\star} \int_0^{\tau_\star} L_{\rm total}(t) dt, 
\end{equation}
which includes the feedback luminosity from accretion flows (Eq. \ref{eq:ltotal}). 
This quantity is computed from the \texttt{MESA} models for a) the entire main sequence, 
including the initial ramp up to $M_{\rm max}$ (${\tilde L}_{\rm 1}$ in phase 1); 
b) the duration of main sequence between $M_\star=M_{\rm max}$ to the 
terminal age main sequence TAMS with 
$M_{\rm max} \geq M_\star \geq M_{\rm TAMS}$ (${\tilde L}_{2}$ in phase 2);
He ignition to depletion (${\tilde L}_{\rm 3}$ in phase 3); and
C ignition to core collapse (${\tilde L}_{\rm 4}$ in phase 4).  

The last two columns in Table~\ref{tab:grid_mixing_all} indicate that 
${\tilde L}_{\rm total}$ in phase 2 (${\tilde L}_{2}$) is much larger 
than that including entire MS (phase 1+2, ${\tilde L}_{\rm 1+2}$), especially for relatively small 
$\rho_{\rm c}$.  This difference is caused by the slow initial (when $M_\star 
\sim 1 M_\odot$) Bondi accretion.  But this difference is modest ($\sim 2-3$) in 
the high-$\rho_{\rm c}$
limit, when ${\dot M}_{\rm acc}$ is limited by ${\dot M}_{\rm acc, KE}$ (Eq. \ref{eq:acclimit}).
In general, $L_{\rm IMS} \gtrsim 2 {\tilde L}_2$. 
Since $\Upsilon (M_\star, Y_{\star})$ is a decreasing function of $\lambda_0$, 
$M_{\rm max}$ (with $\Upsilon_{\rm equi}$ in Eq. \ref{eq:upsiloneq}) increases with
$\lambda_0$, i.e. less efficient feedback generally leads to larger 
${\tilde L}_{\rm total}$ and its corresponding average mass. 
Moreover, ${\tilde L}_2$ is several times larger than that during the 
post main–sequence evolution (${\tilde L}_{3+4}$) 
because a) $M_\star$ has 
already reduced substantially on the main sequence and b) all metamorphic stars 
have approximately Eddington-limited luminosity.

\subsubsection{Implication on stellar surface density and formation rate}

In an opaque disk, radiative diffusion leads to a surface cooling rate of 
\begin{equation}
    \mathcal{Q}^- = {32 \sigma T_c^4 \over 3 \kappa \Sigma} = {8 c \rho c_{\rm s, rad}^2 \over \kappa \Sigma},
\end{equation}
Assuming the embedded metamorphic stars have a uniform age distribution and all their
luminosity is converted into thermal energy of the disk gas, their heating rate per
unit are would be $\mathcal{Q}^+ _\star \simeq s_\star {\tilde L}_{\rm total}$.  In a thermal 
equilibrium ($\mathcal{Q}^+ _\star = \mathcal{Q}^-$), the surface density of stars would be \citep{ChenLin2024}

\begin{equation}
    s_\star = {\mathcal{Q}^- \over {\tilde L}_{\rm total}}= {4 c \over \kappa} {\Omega c_{\rm s, rad}
    \over {\tilde L}_{\rm total}}.
\label{eq:sstar}
\end{equation}

Since the metamorphic stars are generally less luminous that the immortal stars
(Fig. \ref{fig:mass-growth}), their ${\tilde L}_{\rm total}$ is also less than 
$L_\star$ of the immortal stars.  Consequently,  more metamorphic stars than 
immortal stars are needed to maintain the thermal equilibrium
of disks with marginal gravitational stability (with Toomre $Q \simeq 1$).  

In a quasi accretion–wind equilibrium with $\lambda_\star \sim \lambda_0$, 
${\tilde L}_{\rm total}$ also provide an estimate on the average mass of the star
${\tilde M}_\star \sim M_\odot {\tilde L}_{\rm total}/ \lambda_0 L_{\rm Edd, \odot}$. 
Although the metamorphic stars' average mass is a fraction of its maximum value 
($\sim M_{\rm max} (\lambda_0) (1-Y_{\rm d})^2 \sim M_{\rm IMS}$ in Eq. \ref{eq:mstarlambday}), 
their total mass surface density $s_\star {\tilde M}_\star$ and flux are the same as the 
immortal stars.

All metamorphic stars loss mass to the disk during their 
evolution.  
As they fade, the disk contracts with smaller $Q$
and resumption of gravitational instability and fragmentation
\citep{Chen2023}.  Over time, the thermal equilibrium of 
the disk is restored and maintained with the formation of 
a new generation of metamorphic stars at a rate
${\dot s}_\star \sim s_\star/\tau_\star$.

The newly formed stars' initial growth timescale ($\tau_{\rm acc} 
= M_\star / \dot{M}_{\rm acc}$; Eq.~\ref{eq:acclimit}) depends on 
$\rho_{\rm c}$ (Eq. \ref{eq:mdotbondi}) and $M_\star$ (Eq. 
\ref{eq:taueddke}) which may have an initial range extending to 
the sub-solar limit.  Moreover, $\tau_{\rm Bondi} \gg 
\tau_{\rm dyn} = \Omega^{-1}$ for low-mass ($M_\star 
\lesssim 10 M_\odot$) stars (\S\ref{sec:detailedrad}).  
This bottleneck leads to an over-production of low-mass stars which 
subsequently merge and grow much more rapidly than through gas 
accretion (Wang et al. in prep).  In a sufficiently 
populated stellar system, the mass of all continually forming 
metamorphic stars can rapidly grow to $\sim M_{\rm max} 
(\lambda_0) (1-Y_{\rm d})$ and thereafter a quasi-equilibrium can 
be established in which accretion and wind mass-loss balance 
($\dot{M}_{\rm net} \simeq 0$, or $\dot{M}_{\rm acc} \simeq \dot{M}_{\rm wind}$ 
in Eq.~\ref{eq:mdotnet}).  This balance enables self-regulation of the 
star formation rate in the disk and prevents runaway feedback instabilities.

\subsection{Chemical Yields without Wind Retention}
\label{sec:chemyield}

In this section, we discuss the net elemental yield from
both immortal and metamorphic stars to the disk. 
We integrate the net yield (mass return from star to the disk) for element 
Z (including C, N, and O separately)

\begin{equation}
    \Delta M_{\rm Z} 
    =  \int_{\Delta \tau} ({\dot M}_{\rm wind} Z_\star^\prime - {\dot M}_{\rm acc} Z_{\rm d}) dt
\label{eq:deltamz}
\end{equation}

where $Z_\star ^\prime$ is the abundance near the mass losing layer on the stellar 
surface and $\Delta \tau$
is the time interval for relevant stages of stellar evolution.
Note that $Z_\star^\prime$, ${\dot M}_{\rm wind}$, and ${\dot M}_{\rm acc}$ are all functions
of $M_\star$. The He return is calculated with a similar expression.

During phase 1, $M_\star$ ramps up with ${\dot M}_{\rm acc} \gg {\dot M}_{\rm wind}$,
so that the retention effect is negligible and $\Delta M_{\rm Z} \sim - M_{\rm max} 
Z_{\rm d}$. But, stellar wind steeply intensifies as $M_\star \sim  M_{\rm max}$ 
with $\lambda_\star \sim \lambda_0$ (\S\ref{sec:accimplementation}). 
In this subsection, we consider default models that all the gas released by the stellar wind is completely 
return to the disk, 
i.e. without any retention.

\subsubsection{C+O depletion by Immortal stars}
\label{sec:cplusodeplete}

If the freshly-accreted gas can
mix with gas in the convective core, the replenishment of H–fuel to
would indefinitely prolong the main–sequence evolution and render 
the stars embedded in AGN disks ``immortal''.  Through the CNO cycle, 
these stars convert H into He at a rate ${\dot M}_{\rm He} 
= L_\star (\lambda_0) / \epsilon_{\rm He} c^2$.  Concurrently, 
it also converts C and O into N at a rate of ${\dot M}_{\rm N}$
while the total C+N+O abundance is conserved. 

\begin{figure*}[p!]
\gridline{
\fig{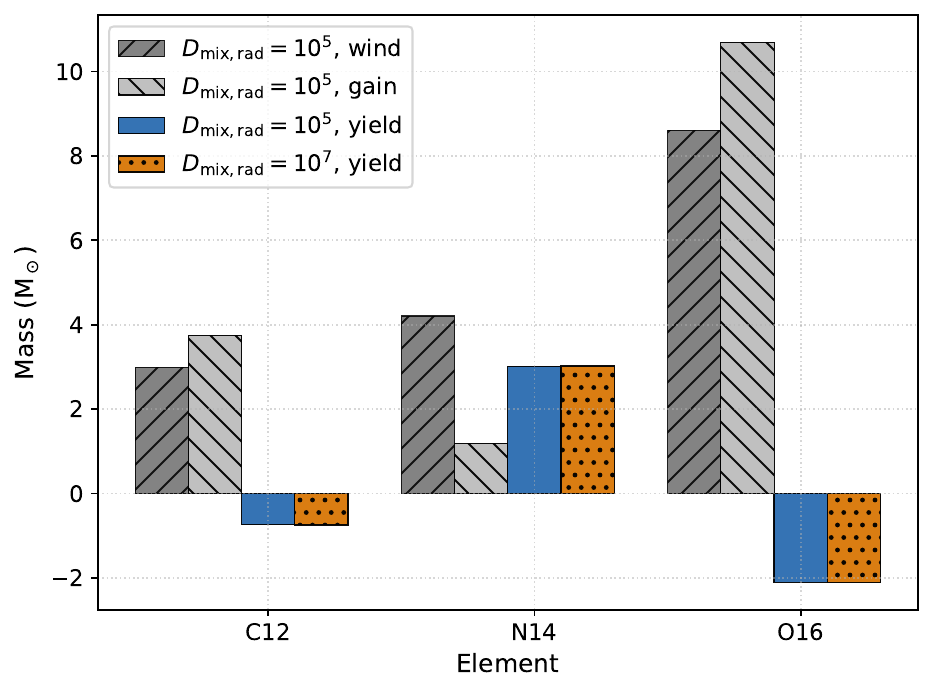}{0.45\textwidth}{(a) Metamorphic star, CNO yields, init $\rightarrow$ central H depletion}
  \fig{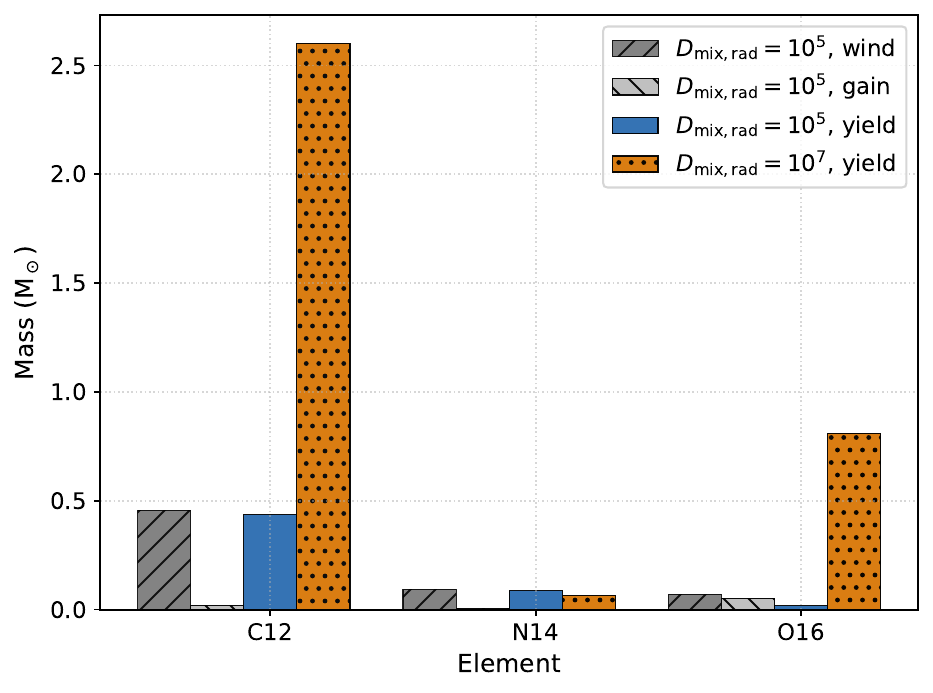}{0.45\textwidth}{(b) Metamorphic star, CNO yields, central H depletion $\rightarrow$ central He depletion}
}
\vspace{-0.2em}
\gridline{
  \fig{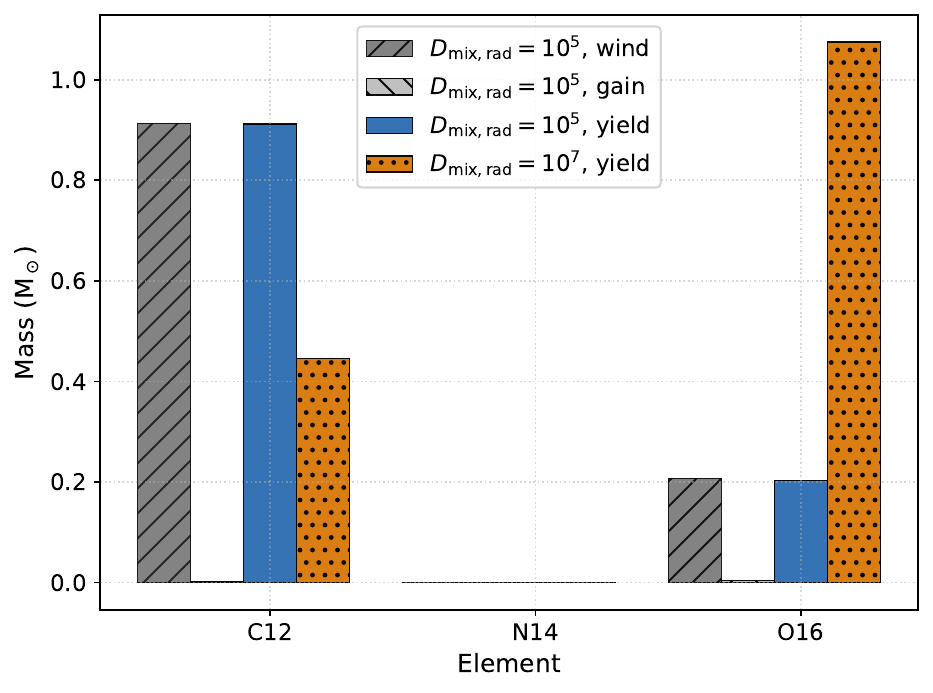}{0.45\textwidth}{(c) Metamorphic star, CNO yields, central He depletion $\rightarrow$ pre-collapse}
  \fig{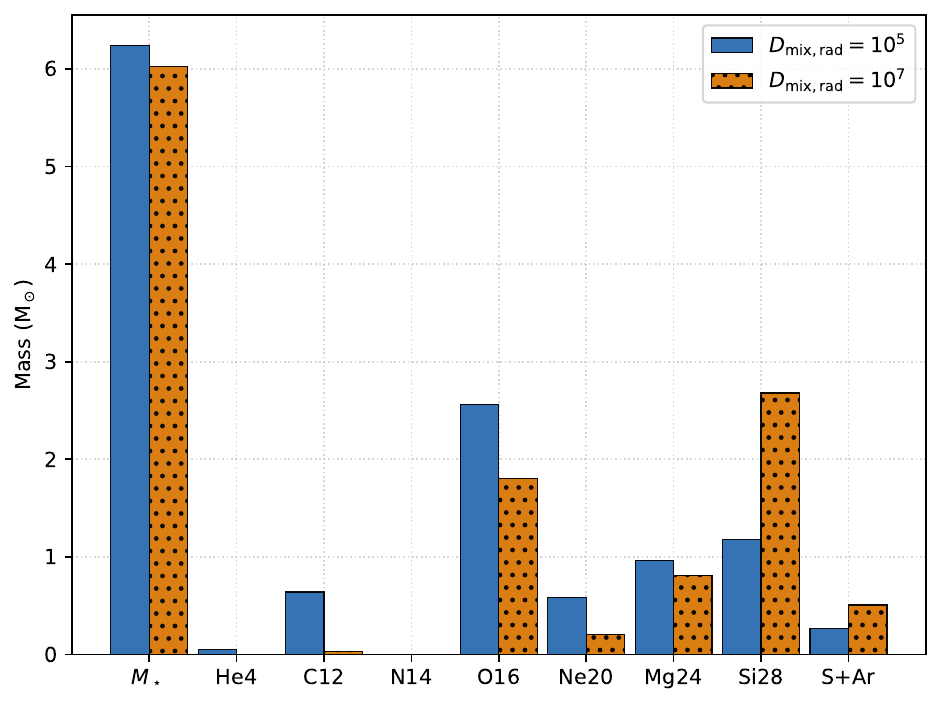}{0.45\textwidth}{(d) Metamorphic star, pre-collapse composition totals}
}
\vspace{-0.2em}
\gridline{
\fig{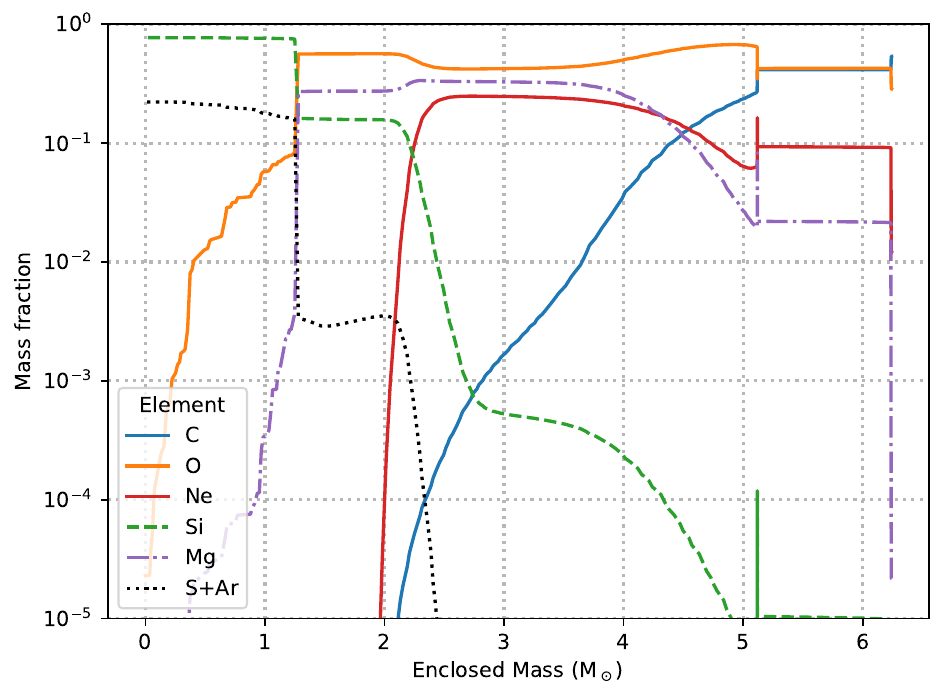}{0.45\textwidth}{(e) Metamorphic star with $D_{\rm mix,rad}=10^5$, pre-collapse chemical radial distribution}
\fig{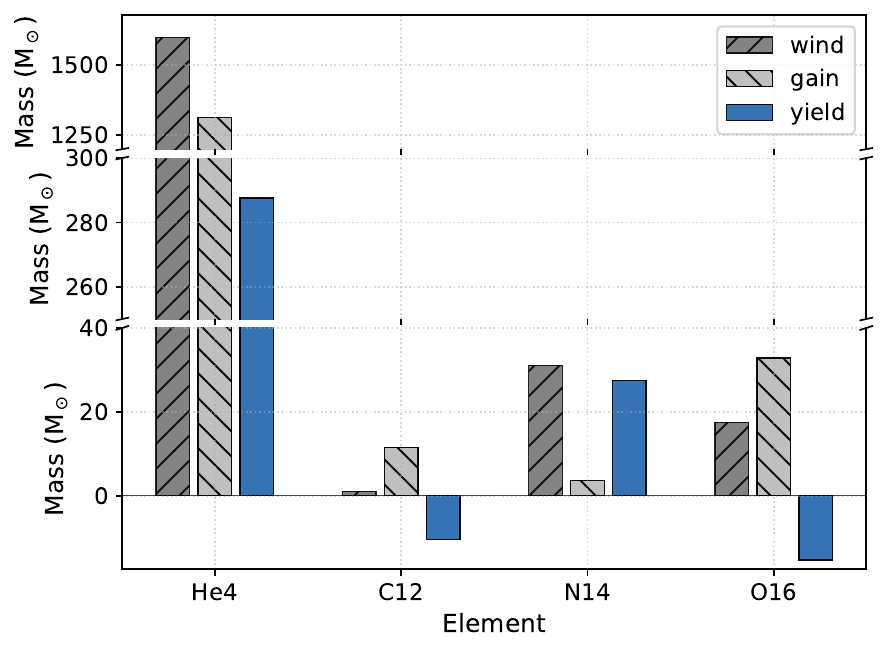}{0.45\textwidth}{(f) Immortal star, He+CNO yields, peak mass $\rightarrow$ model end}
}
\vspace{-0.2em}
  \caption{Metamorphic star's
  cumulative C, N, and O yields (wind minus gain) with 
  $D_{\mathrm{mix, rad}}=10^7\,\mathrm{cm^2\,s^{-1}}$ (red) vs
  $D_{\mathrm{mix, rad}}=10^5\,\mathrm{cm^2\,s^{-1}}$ (green) during
  pre and main–sequence phases 1+2 (top left), He-burning phase 3 (top right), 
  and carbon-burning phase 4 (middle left).  
  Pre-collapse mass of various elements and 
  compositional stratification inside a metamorphic star are 
  shown in the middle-right and lower-left panels respectively. 
  Immortal star's steady He and N yield versus C and O drain,
  accumulated over $\approx$5 Myrs, are shown in the bottom-right panel.
  All stellar models use $Y_{\rm d}=0.25$, $\lambda_0=0.75$, 
  and $\rho_{\rm c}=10^{-16}\,\mathrm{g\,cm^{-3}}$.}
  \label{fig:yields}
\end{figure*}

For a representative model with $\lambda_0=0.75$ and 
$\rho_{\rm c} = 10^{-16} {\rm g \ cm^{-3}}$, we estimate 
${\dot M}_{\rm He} \sim 5 \times 10^{-5} M_\odot {\rm yr^{-1}}$  and
${\dot M}_{\rm N} \sim 5 \times 10^{-6} M_\odot {\rm yr^{-1}} \sim 0.1 {\dot M}_{\rm He}$
(bottom-right, panel (f), Fig. \ref{fig:yields}).  Since the fusion byproducts are
recycled, the disk is polluted in He and N at rates ${\dot M}_{\rm He}$
and ${\dot M}_{\rm N}$ per star respectively.  The conservation of 
C+N+O also implies that the C+O are depleted at a rate 
$\sim 0.5 {\dot M}_{\rm N}$.
Composite quasar spectra and photo‐ionization modeling  
show modest enhancement in He.  The median BLR metallicity of 
$Z\approx4$–$6\,Z_\odot$ for all $\alpha$-elements (including C, 
N, and O) across $2\lesssim z\lesssim6$, without a significant 
redshift evolution (Fig. 6 in \citet{Huang2023}).
These spectroscopic data
is not consistent with the prolific return of He to and 
depletion of C and O inferred for the immortal stars.

\subsubsection{Metamorphic stars' Main sequence yields}
\label{sec:nenrichment}
In contrast, the radiative layer of metamorphic stars separates their core from the surface 
region where gas is accreted and return to the disk.  As He and N are enhanced with the 
depletion of H, C, and O, they are well mixed in the core.  As these 
stars loss mass, the boundary between the convective core and radiative
envelope rezones. The N-laden gas later exudes into the surface layers 
and is eventually released into the surrounding disk.

With the solar values of $Y_{\rm d}$ and $Z_{\rm d}$, we find the N yields to be
$\simeq1.4\,M_\odot$ for $\lambda_0=0.50$,
$\simeq3.1\,M_\odot$ for $\lambda_0=0.75$, and $\simeq4.5\,M_\odot$ for $\lambda_0=0.90$
during phase 1 and 2 (top panel, Fig.~\ref{fig:yields}).
In relative terms this main–sequence period 
spans 6 Myr,  so a single metamorphic 
star can seed its local zone with more than
$0.2-0.5 \rm{M_\odot}$ N on a Myr timescale.
Since C+N+O is conserved in the CNO cycle,
the positive N yield is accompanied by negative C and O yields during the main
sequence evolution, similar to the immortal stars
(top panel, Fig. \ref{fig:yields}).

\subsubsection{Post-main–sequence yields}
\label{sec:postmainsequenceyield}
However, H–exhaustion in the core leads to the transition 
to post-main–sequence evolution with He ignition,
and the production of C and O in the phase 3
(middle-left panel, Fig. \ref{fig:yields}). 
The average mass loss rate $\dot M\sim10^{-5}$–$10^{-4}\,M_\odot\,$yr$^{-1}$
(upper panel Fig. \ref{fig:mdot-lledd-age}). 
Since the stellar envelope is near-Eddington, even
increasing nuclear luminosity from the center,
the star doesn't impulsively expand into super giants.
Instead, it breaks through the outer shells, 
delivering the bulk of He, H and 
other elements to the surroundings.  
{The C and O yield depends on $D_{\rm mix,rad}$.
with the nominal $10^5\,\mathrm{cm^2\,s^{-1}}$ (\S\ref{sec:mixing}),
limited diffusion in the outer envelope 
returns less C and O during phases 3–4 
than were consumed in phases 1–2 (Fig.~\ref{fig:yields}).
}

During phase 3, $\Delta R$ of the radiative layer is smaller and 
the triple-$\alpha$ reaction converts He into $\alpha$-elements on 
shorter timescale that $\tau_{\rm He}$.  Comparison of yield between 
models with $D_{\rm mix, rad} = 10^5$ cm$^2$ s$^{-1}$ and $10^7$ cm$^2$ 
s$^{-1}$ indicate that a) the $\alpha$-element yield increase with 
$D_{\rm mix, rad}$ and b) the latter is adequate for the $\alpha$-element 
byproducts in the core to diffuse through the radiative layer, be carried 
away by the wind and contribute to the stellar yield to the disk.
But, in the limit $ D_{\rm mix, rad} \sim 10^{9-10}$ cm$^2$ s$^{-1}$, 
there is sufficient mixing for the newly-accreted H–laden 
disk gas to prolong the shell burning (via CNO-cycle), despite modest 
${\dot M}_{\rm acc} (\sim 10^{-7} M_\odot$ yr$^{-1}$), 
to sustain ``immortal'' post-main–sequence 
evolution with small residual $M_\star (\lesssim 3 M_\odot)$ and thereby 
prevent the onset of C burning in the core. 

During phase 4, the $\alpha$-chain reaction converts C and O 
into Mg and Si byproducts in the core (middle-right panel of 
Fig \ref{fig:yields}). For $\lambda_0=0.75$ the combined 
C+O mass fraction drops from $90.78\%$ at He depletion to $55.1\%$ 
just prior to collapse, while for $\lambda_0=0.90$ it falls from 
$92.9\%$ to $47.5\%$.  The late-stage CO consumption both boosts 
the eventual yield of intermediate-mass elements (Ne, Mg, and Si) 
and suppresses the residual C/O ratio.

\subsubsection{Supernova yields.}
\label{sec:supernovayield}

With the standard $D_{\rm mix, rad}$ prescription, 
the \texttt{MESA} models terminate at the onset of Si burning when central temperature approaches $\simeq10^{9.5}$K.
Substantial masses of Ne, Mg, Si, S remain in the residual core
with a surrounding C and O envelope.
The core quickly runs out of nuclear fuel and undergoes collapse into
either a black hole or a neutron star of a few $M_\odot$. By the 
pre-collapse stage, Mg and Si together exceed $\sim$20\% of the 
remaining stellar mass (bottom-left panel (e) Fig.\ref{fig:yields}), 
S and Ar are poised to rise sharply while
Fe‐group nuclei, though still $\lesssim10^{-3}$ by mass before explosive Si burning is ignited.

The stratified composition in the pre-collapse stellar envelope 
(bottom-left panel Fig.\ref{fig:yields}) determines 
the relative abundance of the yield \citep{Fryer2025}.  
With identical $D_{\text {mix }} (=10^5$ cm$^2$ s$^{-1}$), but different $\lambda_0$, the models are similar with each other in the mass-fraction ratios, internal profiles, and overall structure. 

MESA's limited nuclear reaction network posts uncertainties on the pre-collapse composition
and structure in the core beyond the O/Si-burning phase \citep{Renzo2024}. In a follow-up study, we 
will use  comparison models \citep{Fryer2025} with more comprehensive treatment of the 
nuclear reaction network \citep{woosley2002, heger2005}, rotation, magnetic fields, turbulence, 
and neutrino cooling \citep{fryer2018, andrews2020} to examine the sensitivity to the choice of 
reaction networks and diffusion efficiency in MESA.

\subsection{Chemical Yield with Total Retention}
\label{sec:retention1}

We also consider a set of analogous models under the 
assumption that, with its yield, the stellar wind contaminates the 
proximity of the stars which subsequently re-accreted, {\it in situ}, 
the polluted gas (\S\ref{sec:accimplementation}).  This effect is particularly important during
the prolonged main–sequence evolution (phase 2) when both ${\dot M}_{\rm acc}$
and ${\dot M}_{\rm wind}$ are higher than ${\dot M}_{\rm net}$,
i.e. considerable amount of gas is being exchanged 
with or without net changes in $M_\star$
(Fig. \ref{fig:mdot-lledd-age}).  

For stars with ``extra mixing'', this recycle process is equivalent to 
chemical insulation which suppress the replenishment of H–rich disk gas.
\citet{AlidibLin2023} have previously shown and we confirm that nearly total 
($\gtrsim 90\%$) retention quenches fresh H supply from the disk
to the star (including the nuclear burning core) and leads to transition
to metamorphic stars even with extra mixing.  

For metamorphic stars (with conventional mixing), 
$\Delta M_{\rm Z} \sim - M_{\rm max} Z_{\rm d}$ during phase 1 regardless of retention efficiency since the stellar composition has hardly changed.
During $\Delta \tau_2$ of main sequence phase 2, total retention 
is represented by a modified Equation (\ref{eq:deltamz})
$    \Delta M_{\rm Z} 
    =  -\int_{\Delta \tau_2} {\dot M}_{\rm net} Z_\star^\prime dt
    \simeq {\tilde Z}_\star^\prime (M_{\rm max} - M_{\rm collapse})$
where ${\tilde Z}_\star^\prime$ is mass-weighted abundance in the
wind-launching outer region of the star.
Difference in the nuclear burning rates at each step along the CNO cycle
monotonically increases N abundance $Z_{\star, \rm N}$ 
(and decreases C \& O abundance $Z_{\star, \rm C}$ \& $Z_{\star, \rm O}$) 
as $M_\star \rightarrow M_{\rm TAMS}$ with $Y_\star \rightarrow 1$
(Fig.7 Ali-Dib \& Lin, 2023).  Consequently,
\begin{equation}
    Z_{\star, \rm N} ^\prime (M_{\rm max}) \leq {\tilde Z}_\star^\prime \leq Z_{\star, \rm N} ^\prime (M_{\rm TAMS})
\end{equation}
and
\begin{equation}
Z_{\star, \rm C}^\prime (M_{\rm max}) \geq {\tilde Z}_\star^\prime \geq Z_{\star, \rm C} ^\prime (M_{\rm TAMS}) \text{(also for O).}
\end{equation}

Since strong stellar wind dominates diminishing accretion during the 
post main–sequence evolution (phases 3 and 4 in the upper-right and middle-left panels 
of Fig. \ref{fig:yields}), such that retention does not significantly 
contribute to the star's internal composition and 
$\Delta M_{\rm Z} \simeq {\tilde Z}_{\star, 3+4}^\prime 
(M_{\rm TAMS} - M_{\rm collapse})$ and ${\tilde Z}_{\star, 3+4}^\prime$
is mass averaged over phases 3 and 4.  Since $M_{\rm max} \gg M_{\rm TAMS}$ 
and $M_{\rm collapse}$ (Fig \ref{fig:mass-growth}) $\Delta M_{\rm Z}
\simeq ({\tilde Z}_{\star}^\prime - Z_{\rm d}) M_{\rm max}$. 

We adopt $\rho_{\rm c} = 10^{-16}\,\mathrm{g\,cm^{-3}}$ and $\lambda_0=0.75$ which
gives $M_{\rm max}=234 M_\odot$ and $M_{\rm TAMS} = 14.75 M_\odot$.
Even if all the C and O mass (based on the disk abundance \S\ref{sec:parameters})
are converted into N during phase 2, the net yield would be 
$\Delta M_{\rm Z_N} \sim 1.87 M_\odot$ for N, $\Delta M_{\rm Z_C} \sim - 0.49 M_\odot$ for C, 
and $\Delta M_{\rm Z_O} \sim -1.38 M_\odot$.  The magnitudes of these quantities are more than half
those provided by the zero-retention \texttt{MESA} models (lower-right Fig. \ref{fig:yields}).
Since the yield during phases 1, 3, and 4 do not dependent on the retention efficiency,
these estimates provide fractionally more positive total C, N, and O yield.

\section{Summary and Discussions}\label{sec:sum}
Conventional $\alpha$-model implies AGN disks become
gravitationally unstable outside $\sim 10^{2-3} R_\bullet$.
Subsequent fragmentation leads to spontaneous {\it in situ} 
star formation. Throughout their lifespan, these stars 
reside in gaseous environment with densities many orders of 
magnitude greater than that of dense molecular cloud cores.
In contrast to the stand-alone massive stars in the field,
the embedded stars rapidly gain mass at rates which are
self-regulated by the radiative feedback initially from the 
dissipated accretion energy and subsequently from their
intrinsic nucleosynthesis.  
When they become sufficiently
massive, their intrinsic luminosity approaches its Eddington 
limit, quenches accretion, and drives intense winds to 
halt further growth.  

The limiting stellar mass {($M_{\rm max} \sim$ several tens to hundreds of solar masses)} for the 
onset of this quasi accretion–wind equilibrium increases with the 
feedback-efficiency factor $\lambda_0$ and decreases with helium 
abundance of the disk gas $Y_{\rm d}$.  But it is insensitive to the 
background density $\rho_{\rm c}$ and sound speed $c_{\rm s}$.

Accretion and wind exchange gas between the disk and stars' 
surface layer which is separated from their nuclear-burning convective 
core.  While the diffusion coefficient in the convective core has been 
conventionally estimated using the mixing length model, that of 
the radiative envelope range from minute molecular values to 
{\it ad hoc} prescriptions for ``extra mixing'' associated with 
potential rotational circulation or radiative feedback.  
In view of this uncertain, we presented a series of \texttt{MESA} models
with a range of mixing efficiency in the radiative envelope.

We show the magnitude of $D_{\rm mix, rad}$ essentially 
determines the pathways of subsequent evolution \citep{Xu2025a}. With an assumed extra mixing for the radiative 
envelope, the continuous replenishment of H into the core enables 
immortal stars to produce and release  He and N yields 
with unceasing drainage of C and O from the disk.  
This expectation is not consistent with the observed super-solar 
abundance enhancement for all $\alpha$-element (including 
C, N, and O), independent of redshift \citep{Huang2023}.  

With a conventional diffusivity prescription, the He ashes
accumulates in insulated core of main–sequence metamorphic star. 
These stars shed mass to maintain nearly Eddington limited 
luminosity and a quasi accretion–wind equilibrium.
Concurrent accretion and stellar wind also lead to N yield 
to and C+O removal from the disk.  During the subsequent 
post-main–sequence evolution, triple-$\alpha$ process 
leads to the conversion of He into light $\alpha$ elements (mostly C 
and O) with a positive yields to the disk. In the late stages of 
stellar evolution, $\alpha$-chain reaction converts the light 
into heavy $\alpha$ elements with
chemically stratified stellar structure.  The metamorphic stars
undergo supernova and return most of the residual C and O in the
envelope as well as some Mg and Si in the core and Fe produced
in circum-stellar disks around their compact remnants. Metamorphic stars' 
prolific and robust produce of heavy elements are in general
agreement with the observed super-solar $\alpha$-elements and 
sub-solar Fe abundances of AGN BLRs \citep{Huang2023}.
The differential magnitude of individual
elements' yield depends primarily the diffusivity during 
the post main–sequence evolution and star's pre-collapse
angular momentum distribution.

Based on their evolution track, we extrapolate the
metamorphic stars' mass function and surface density $s_\star$ 
required to maintain a state of marginal gravitational 
(in)stability for the AGN disk.  Metamorphic stars continually
form at a self-regulated rate as their evolve (on a few Myr 
timescale) into stellar mass (a few $M_\odot$) compact 
remnants. In a follow-up paper, we will infer, 
from AGN's observed slightly subsolar Fe abundance,
the production rate and evolution of stellar-mass black-hole 
population.

This unceasing self-regulated star formation efficiency 
fundamentally differs from that of stand-along star forming
regions in the Galaxy.  Moreover, the heavy-element yields
are deposited into the AGN disks and join the accretion flow 
toward the central supermassive black holes. Since
these pollutants do not accumulate over time, the metallicity
of the BLRs reach a saturation level which is independent
of the cosmic redshift \citep{Huang2023}.  This mechanism 
differs from the traditional explanation of high quasar metallicities 
(which often invokes an earlier phase of starburst and enrichment in 
the host galaxy) by placing the enrichment process {\it in situ} within 
the AGN–disk.  Detailed analysis
on the chemical evolution of AGN disks will be presented 
in subsequent works.


Finally, it is worth noting that in the very dense inner regions of the disk where $\tau_{\rm merge} 
\lesssim \tau_{\rm wind}$ (Eq. \ref{eq:tauwind}), stars may coalesce 
before they can re-establish a quasi hydrostatic equilibrium, undergo 
runaway growth beyond $10^3 M_\odot$.
The embedded very-massive stars (VMS) scenario also has implication for 
transient phenomena in AGNs. For instance, if one of these massive stars 
undergoes a supernova, it would occur deeply embedded in the AGN–disk. 
The interaction of supernova ejecta with the dense disk gas could lead 
to a shock breakout signature or a luminous radio afterglow observable 
in conjunction with an AGN. There has been speculation that some unusual 
transient events in galactic nuclei (sometimes labeled as ``changing-look AGN" or peculiar nuclear flares) might be attributed to stellar transients within disks. Our study provides a concrete model for one class of such events.


\section*{Data Availability}
All simulation data, inlists, custom routines, and output used in this work are available on Zenodo \citep{Xu2025Zenodo} and
\url{https://github.com/zhxu-astro/AGNstarRadFB}.

\section*{Acknowledgments}
The authors thank Mohamad Ali-Dib, Chris Fryer, Stan Woosley, Long Wang, Jiamu Huang, 
Sunny Wong, Keith Horne, 
Jeremy Goodman, and Yan-Fei Jiang for useful conversations.
We thank the Institute of Advanced Studies Tsinghua University
and the Department of Astronomy Westlake University for hospitality and support.  

\bibliography{main}{}

\begin{thebibliography}{}
\expandafter\ifx\csname natexlab\endcsname\relax\def\natexlab#1{#1}\fi
\providecommand{\url}[1]{\href{#1}{#1}}
\providecommand{\dodoi}[1]{doi:~\href{http://doi.org/#1}{\nolinkurl{#1}}}
\providecommand{\doeprint}[1]{\href{http://ascl.net/#1}{\nolinkurl{http://ascl.net/#1}}}
\providecommand{\doarXiv}[1]{\href{https://arxiv.org/abs/#1}{\nolinkurl{https://arxiv.org/abs/#1}}}

\bibitem[{{Ali-Dib} \& {Lin}(2023)}]{AlidibLin2023}
{Ali-Dib}, M., \& {Lin}, D. N.~C. 2023, \mnras, 526, 5824,
  \dodoi{10.1093/mnras/stad2774}

\bibitem[{{Andrews} {et~al.}(2020){Andrews}, {Fryer}, {Even}, {Jones}, \&
  {Pignatari}}]{andrews2020}
{Andrews}, S., {Fryer}, C., {Even}, W., {Jones}, S., \& {Pignatari}, M. 2020,
  \apj, 890, 35, \dodoi{10.3847/1538-4357/ab64f8}

\bibitem[{{Angulo} {et~al.}(1999){Angulo}, {Arnould}, {Rayet}, {Descouvemont},
  {Baye}, {Leclercq-Willain}, {Coc}, {Barhoumi}, {Aguer}, {Rolfs}, {Kunz},
  {Hammer}, {Mayer}, {Paradellis}, {Kossionides}, {Chronidou}, {Spyrou},
  {degl'Innocenti}, {Fiorentini}, {Ricci}, {Zavatarelli}, {Providencia},
  {Wolters}, {Soares}, {Grama}, {Rahighi}, {Shotter}, \& {Lamehi
  Rachti}}]{Angulo1999}
{Angulo}, C., {Arnould}, M., {Rayet}, M., {et~al.} 1999, \nphysa, 656, 3,
  \dodoi{10.1016/S0375-9474(99)00030-5}

\bibitem[{{Artymowicz} {et~al.}(1993){Artymowicz}, {Lin}, \&
  {Wampler}}]{Artymowicz1993}
{Artymowicz}, P., {Lin}, D.~N.~C., \& {Wampler}, E.~J. 1993, \apj, 409, 592,
  \dodoi{10.1086/172690}

\bibitem[{{Blouin} {et~al.}(2020){Blouin}, {Shaffer}, {Saumon}, \&
  {Starrett}}]{Blouin2020}
{Blouin}, S., {Shaffer}, N.~R., {Saumon}, D., \& {Starrett}, C.~E. 2020, \apj,
  899, 46, \dodoi{10.3847/1538-4357/ab9e75}

\bibitem[{{Cantiello} {et~al.}(2021){Cantiello}, {Jermyn}, \&
  {Lin}}]{Cantiello2021}
{Cantiello}, M., {Jermyn}, A.~S., \& {Lin}, D. N.~C. 2021, \apj, 910, 94,
  \dodoi{10.3847/1538-4357/abdf4f}

\bibitem[{{Cassisi} {et~al.}(2007){Cassisi}, {Potekhin}, {Pietrinferni},
  {Catelan}, \& {Salaris}}]{Cassisi2007}
{Cassisi}, S., {Potekhin}, A.~Y., {Pietrinferni}, A., {Catelan}, M., \&
  {Salaris}, M. 2007, \apj, 661, 1094, \dodoi{10.1086/516819}

\bibitem[{{Chen} {et~al.}(2025){Chen}, {Jiang}, \& {Goodman}}]{Chen2025}
{Chen}, Y.-X., {Jiang}, Y.-F., \& {Goodman}, J. 2025, arXiv e-prints,
  arXiv:2505.13951, \dodoi{10.48550/arXiv.2505.13951}

\bibitem[{{Chen} {et~al.}(2024){Chen}, {Jiang}, {Goodman}, \& {Lin}}]{Chen2024}
{Chen}, Y.-X., {Jiang}, Y.-F., {Goodman}, J., \& {Lin}, D. N.~C. 2024, \apj,
  974, 106, \dodoi{10.3847/1538-4357/ad6dd4}

\bibitem[{{Chen} {et~al.}(2023){Chen}, {Jiang}, {Goodman}, \&
  {Ostriker}}]{Chen2023}
{Chen}, Y.-X., {Jiang}, Y.-F., {Goodman}, J., \& {Ostriker}, E.~C. 2023, \apj,
  948, 120, \dodoi{10.3847/1538-4357/acc023}

\bibitem[{{Chen} \& {Lin}(2024)}]{ChenLin2024}
{Chen}, Y.-X., \& {Lin}, D. N.~C. 2024, \apj, 967, 88,
  \dodoi{10.3847/1538-4357/ad3c3a}

\bibitem[{{Chugunov} {et~al.}(2007){Chugunov}, {Dewitt}, \&
  {Yakovlev}}]{Chugunov2007}
{Chugunov}, A.~I., {Dewitt}, H.~E., \& {Yakovlev}, D.~G. 2007, \prd, 76,
  025028, \dodoi{10.1103/PhysRevD.76.025028}

\bibitem[{{Cyburt} {et~al.}(2010){Cyburt}, {Amthor}, {Ferguson}, {Meisel},
  {Smith}, {Warren}, {Heger}, {Hoffman}, {Rauscher}, {Sakharuk}, {Schatz},
  {Thielemann}, \& {Wiescher}}]{Cyburt2010}
{Cyburt}, R.~H., {Amthor}, A.~M., {Ferguson}, R., {et~al.} 2010, \apjs, 189,
  240, \dodoi{10.1088/0067-0049/189/1/240}

\bibitem[{{Dittmann} {et~al.}(2021){Dittmann}, {Cantiello}, \&
  {Jermyn}}]{Dittmann2021}
{Dittmann}, A.~J., {Cantiello}, M., \& {Jermyn}, A.~S. 2021, \apj, 916, 48,
  \dodoi{10.3847/1538-4357/ac042c}

\bibitem[{{Fabj} {et~al.}(2025){Fabj}, {Dittmann}, {Cantiello}, {Perna}, \&
  {Samsing}}]{Fabj2025}
{Fabj}, G., {Dittmann}, A.~J., {Cantiello}, M., {Perna}, R., \& {Samsing}, J.
  2025, \apj, 981, 16, \dodoi{10.3847/1538-4357/ada896}

\bibitem[{{Ferguson} {et~al.}(2005){Ferguson}, {Alexander}, {Allard}, {Barman},
  {Bodnarik}, {Hauschildt}, {Heffner-Wong}, \& {Tamanai}}]{Ferguson2005}
{Ferguson}, J.~W., {Alexander}, D.~R., {Allard}, F., {et~al.} 2005, \apj, 623,
  585, \dodoi{10.1086/428642}

\bibitem[{{Fryer} {et~al.}(2018){Fryer}, {Andrews}, {Even}, {Heger}, \&
  {Safi-Harb}}]{fryer2018}
{Fryer}, C.~L., {Andrews}, S., {Even}, W., {Heger}, A., \& {Safi-Harb}, S.
  2018, \apj, 856, 63, \dodoi{10.3847/1538-4357/aaaf6f}

\bibitem[{{Fryer} {et~al.}(2025){Fryer}, {Huang}, {Ali-Dib}, {Andrews}, {Xu},
  \& {Lin}}]{Fryer2025}
{Fryer}, C.~L., {Huang}, J., {Ali-Dib}, M., {et~al.} 2025, \mnras, 537, 1556,
  \dodoi{10.1093/mnras/staf130}

\bibitem[{{Fuller} {et~al.}(1985){Fuller}, {Fowler}, \& {Newman}}]{Fuller1985}
{Fuller}, G.~M., {Fowler}, W.~A., \& {Newman}, M.~J. 1985, \apj, 293, 1,
  \dodoi{10.1086/163208}

\bibitem[{{Goodman} \& {Tan}(2004)}]{GoodmanTan2004}
{Goodman}, J., \& {Tan}, J.~C. 2004, \apj, 608, 108, \dodoi{10.1086/386360}

\bibitem[{{Graham} {et~al.}(2020){Graham}, {Ford}, {McKernan}, {Ross}, {Stern},
  {Burdge}, {Coughlin}, {Djorgovski}, {Drake}, {Duev}, {Kasliwal}, {Mahabal},
  {van Velzen}, {Belecki}, {Bellm}, {Burruss}, {Cenko}, {Cunningham}, {Helou},
  {Kulkarni}, {Masci}, {Prince}, {Reiley}, {Rodriguez}, {Rusholme}, {Smith}, \&
  {Soumagnac}}]{graham2020}
{Graham}, M.~J., {Ford}, K.~E.~S., {McKernan}, B., {et~al.} 2020, \prl, 124,
  251102, \dodoi{10.1103/PhysRevLett.124.251102}

\bibitem[{{Hamann} \& {Ferland}(1999)}]{Hamann1999}
{Hamann}, F., \& {Ferland}, G. 1999, \araa, 37, 487,
  \dodoi{10.1146/annurev.astro.37.1.487}

\bibitem[{{Hamann} {et~al.}(2002){Hamann}, {Korista}, {Ferland}, {Warner}, \&
  {Baldwin}}]{Hamann2002}
{Hamann}, F., {Korista}, K.~T., {Ferland}, G.~J., {Warner}, C., \& {Baldwin},
  J. 2002, \apj, 564, 592, \dodoi{10.1086/324289}

\bibitem[{{Heger} {et~al.}(2005){Heger}, {Woosley}, \& {Spruit}}]{heger2005}
{Heger}, A., {Woosley}, S.~E., \& {Spruit}, H.~C. 2005, \apj, 626, 350,
  \dodoi{10.1086/429868}

\bibitem[{{Huang} {et~al.}(2023){Huang}, {Lin}, \& {Shields}}]{Huang2023}
{Huang}, J., {Lin}, D. N.~C., \& {Shields}, G. 2023, \mnras, 525, 5702,
  \dodoi{10.1093/mnras/stad2642}

\bibitem[{{Iglesias} \& {Rogers}(1996)}]{Iglesias1996}
{Iglesias}, C.~A., \& {Rogers}, F.~J. 1996, \apj, 464, 943,
  \dodoi{10.1086/177381}

\bibitem[{{Itoh} {et~al.}(1996){Itoh}, {Hayashi}, {Nishikawa}, \&
  {Kohyama}}]{Itoh1996}
{Itoh}, N., {Hayashi}, H., {Nishikawa}, A., \& {Kohyama}, Y. 1996, \apjs, 102,
  411, \dodoi{10.1086/192264}

\bibitem[{{Jermyn} {et~al.}(2022){Jermyn}, {Dittmann}, {McKernan}, {Ford}, \&
  {Cantiello}}]{Jermyn2022}
{Jermyn}, A.~S., {Dittmann}, A.~J., {McKernan}, B., {Ford}, K.~E.~S., \&
  {Cantiello}, M. 2022, \apj, 929, 133, \dodoi{10.3847/1538-4357/ac5d40}

\bibitem[{{Jermyn} {et~al.}(2021){Jermyn}, {Schwab}, {Bauer}, {Timmes}, \&
  {Potekhin}}]{Jermyn2021}
{Jermyn}, A.~S., {Schwab}, J., {Bauer}, E., {Timmes}, F.~X., \& {Potekhin},
  A.~Y. 2021, \apj, 913, 72, \dodoi{10.3847/1538-4357/abf48e}

\bibitem[{{Jermyn} {et~al.}(2023){Jermyn}, {Bauer}, {Schwab}, {Farmer}, {Ball},
  {Bellinger}, {Dotter}, {Joyce}, {Marchant}, {Mombarg}, {Wolf}, {Sunny Wong},
  {Cinquegrana}, {Farrell}, {Smolec}, {Thoul}, {Cantiello}, {Herwig}, {Toloza},
  {Bildsten}, {Townsend}, \& {Timmes}}]{MESA2023Jermyn}
{Jermyn}, A.~S., {Bauer}, E.~B., {Schwab}, J., {et~al.} 2023, \apjs, 265, 15,
  \dodoi{10.3847/1538-4365/acae8d}

\bibitem[{{Jiang} \& {Goodman}(2011)}]{JiangGoodman2011}
{Jiang}, Y.-F., \& {Goodman}, J. 2011, \apj, 730, 45,
  \dodoi{10.1088/0004-637X/730/1/45}

\bibitem[{{Lamers} \& {Cassinelli}(1999)}]{lamers1999}
{Lamers}, H. J.~G.~L.~M., \& {Cassinelli}, J.~P. 1999, {Introduction to Stellar
  Winds}

\bibitem[{{Liu} {et~al.}(2025{\natexlab{a}}){Liu}, {Wang}, {Fu}, \&
  {Ho}}]{Liu2025}
{Liu}, M., {Wang}, L., {Fu}, X., \& {Ho}, L.~C. 2025{\natexlab{a}}, \apj, 978,
  87, \dodoi{10.3847/1538-4357/ad91a9}

\bibitem[{{Liu} {et~al.}(2025{\natexlab{b}}){Liu}, {Wang}, \&
  {Peng}}]{Liu2025b}
{Liu}, M., {Wang}, L., \& {Peng}, P. 2025{\natexlab{b}}, arXiv e-prints,
  arXiv:2505.10524, \dodoi{10.48550/arXiv.2505.10524}

\bibitem[{{MacLeod} \& {Lin}(2020)}]{MacLeodLin2020}
{MacLeod}, M., \& {Lin}, D. N.~C. 2020, \apj, 889, 94,
  \dodoi{10.3847/1538-4357/ab64db}

\bibitem[{{Maeder} \& {Meynet}(2010)}]{Maeder2010}
{Maeder}, A., \& {Meynet}, G. 2010, \nar, 54, 32,
  \dodoi{10.1016/j.newar.2010.09.017}

\bibitem[{{McKernan} {et~al.}(2012){McKernan}, {Ford}, {Lyra}, \&
  {Perets}}]{McKernan2012}
{McKernan}, B., {Ford}, K.~E.~S., {Lyra}, W., \& {Perets}, H.~B. 2012, \mnras,
  425, 460, \dodoi{10.1111/j.1365-2966.2012.21486.x}

\bibitem[{{Nagao} {et~al.}(2006){Nagao}, {Maiolino}, \& {Marconi}}]{Nagao2006}
{Nagao}, T., {Maiolino}, R., \& {Marconi}, A. 2006, \aap, 459, 85,
  \dodoi{10.1051/0004-6361:20065216}

\bibitem[{{Oda} {et~al.}(1994){Oda}, {Hino}, {Muto}, {Takahara}, \&
  {Sato}}]{Oda1994}
{Oda}, T., {Hino}, M., {Muto}, K., {Takahara}, M., \& {Sato}, K. 1994, Atomic
  Data and Nuclear Data Tables, 56, 231, \dodoi{10.1006/adnd.1994.1007}

\bibitem[{{Owocki} \& {Shaviv}(2012)}]{owocki2012}
{Owocki}, S.~P., \& {Shaviv}, N.~J. 2012, in Astrophysics and Space Science
  Library, Vol. 384, Eta Carinae and the Supernova Impostors, ed. K.~{Davidson}
  \& R.~M. {Humphreys}, 275, \dodoi{10.1007/978-1-4614-2275-4_12}

\bibitem[{{Paxton} {et~al.}(2011){Paxton}, {Bildsten}, {Dotter}, {Herwig},
  {Lesaffre}, \& {Timmes}}]{MESA2011}
{Paxton}, B., {Bildsten}, L., {Dotter}, A., {et~al.} 2011, \apjs, 192, 3,
  \dodoi{10.1088/0067-0049/192/1/3}

\bibitem[{{Paxton} {et~al.}(2013){Paxton}, {Cantiello}, {Arras}, {Bildsten},
  {Brown}, {Dotter}, {Mankovich}, {Montgomery}, {Stello}, {Timmes}, \&
  {Townsend}}]{MESA2013}
{Paxton}, B., {Cantiello}, M., {Arras}, P., {et~al.} 2013, \apjs, 208, 4,
  \dodoi{10.1088/0067-0049/208/1/4}

\bibitem[{{Paxton} {et~al.}(2015){Paxton}, {Marchant}, {Schwab}, {Bauer},
  {Bildsten}, {Cantiello}, {Dessart}, {Farmer}, {Hu}, {Langer}, {Townsend},
  {Townsley}, \& {Timmes}}]{MESA2015}
{Paxton}, B., {Marchant}, P., {Schwab}, J., {et~al.} 2015, \apjs, 220, 15,
  \dodoi{10.1088/0067-0049/220/1/15}

\bibitem[{{Paxton} {et~al.}(2019){Paxton}, {Smolec}, {Schwab}, {Gautschy},
  {Bildsten}, {Cantiello}, {Dotter}, {Farmer}, {Goldberg}, {Jermyn}, {Kanbur},
  {Marchant}, {Thoul}, {Townsend}, {Wolf}, {Zhang}, \& {Timmes}}]{MESA2019}
{Paxton}, B., {Smolec}, R., {Schwab}, J., {et~al.} 2019, \apjs, 243, 10,
  \dodoi{10.3847/1538-4365/ab2241}

\bibitem[{{Poutanen}(2017)}]{Poutanen2017}
{Poutanen}, J. 2017, \apj, 835, 119, \dodoi{10.3847/1538-4357/835/2/119}

\bibitem[{{Prat} \& {Lignières}(2014)}]{Prat2014}
{Prat}, V., \& {Lignières}, F. 2014, Astronomy \& Astrophysics, 566, A110,
  \dodoi{10.1051/0004-6361/201423655}

\bibitem[{{Renzo} {et~al.}(2024){Renzo}, {Goldberg}, {Grichener}, {Gottlieb},
  \& {Cantiello}}]{Renzo2024}
{Renzo}, M., {Goldberg}, J.~A., {Grichener}, A., {Gottlieb}, O., \&
  {Cantiello}, M. 2024, Research Notes of the American Astronomical Society, 8,
  152, \dodoi{10.3847/2515-5172/ad530e}

\bibitem[{{Samsing} {et~al.}(2022){Samsing}, {Bartos}, {D'Orazio}, {Haiman},
  {Kocsis}, {Leigh}, {Liu}, {Pessah}, \& {Tagawa}}]{Samsing2022}
{Samsing}, J., {Bartos}, I., {D'Orazio}, D.~J., {et~al.} 2022, \nat, 603, 237,
  \dodoi{10.1038/s41586-021-04333-1}

\bibitem[{{Shankar} {et~al.}(2009){Shankar}, {Weinberg}, \&
  {Miralda-Escud{\'e}}}]{shankar2009}
{Shankar}, F., {Weinberg}, D.~H., \& {Miralda-Escud{\'e}}, J. 2009, \apj, 690,
  20, \dodoi{10.1088/0004-637X/690/1/20}

\bibitem[{{Sirko} \& {Goodman}(2003)}]{SirkoGoodman2003}
{Sirko}, E., \& {Goodman}, J. 2003, \mnras, 341, 501,
  \dodoi{10.1046/j.1365-8711.2003.06431.x}

\bibitem[{{Spruit}(2002)}]{Spruit2002}
{Spruit}, H.~C. 2002, \aap, 381, 923, \dodoi{10.1051/0004-6361:20011465}

\bibitem[{{Tagawa} {et~al.}(2020){Tagawa}, {Haiman}, \& {Kocsis}}]{Tagawa2020a}
{Tagawa}, H., {Haiman}, Z., \& {Kocsis}, B. 2020, \apj, 898, 25,
  \dodoi{10.3847/1538-4357/ab9b8c}

\bibitem[{{Thompson} {et~al.}(2005){Thompson}, {Quataert}, \&
  {Murray}}]{Thompson2005}
{Thompson}, T.~A., {Quataert}, E., \& {Murray}, N. 2005, \apj, 630, 167,
  \dodoi{10.1086/431923}

\bibitem[{{Wang} {et~al.}(2022){Wang}, {Jiang}, {Shen}, {Ho}, {Vestergaard},
  {Ba{\~n}ados}, {Willott}, {Wu}, {Zou}, {Yang}, {Wang}, {Fan}, \&
  {Wu}}]{Wang+2022}
{Wang}, S., {Jiang}, L., {Shen}, Y., {et~al.} 2022, \apj, 925, 121,
  \dodoi{10.3847/1538-4357/ac3a69}

\bibitem[{{Wang} {et~al.}(2024){Wang}, {Zhu}, \& {Lin}}]{Wang2024}
{Wang}, Y., {Zhu}, Z., \& {Lin}, D. N.~C. 2024, \mnras, 528, 4958,
  \dodoi{10.1093/mnras/stae321}

\bibitem[{{Woosley} {et~al.}(2002){Woosley}, {Heger}, \&
  {Weaver}}]{woosley2002}
{Woosley}, S.~E., {Heger}, A., \& {Weaver}, T.~A. 2002, Reviews of Modern
  Physics, 74, 1015, \dodoi{10.1103/RevModPhys.74.1015}

\bibitem[{{Xu} {et~al.}(2018){Xu}, {Bian}, {Shen}, {Zuo}, {Fan}, \&
  {Zhu}}]{Xu2018}
{Xu}, F., {Bian}, F., {Shen}, Y., {et~al.} 2018, \mnras, 480, 345,
  \dodoi{10.1093/mnras/sty1763}

\bibitem[{Xu(2025{\natexlab{a}})}]{Xu2025Zenodo}
Xu, Z. 2025{\natexlab{a}}, Code for ``Stellar Evolution with Radiative Feedback
  in AGN Disks'',  Zenodo, \dodoi{10.5281/zenodo.16413005}

\bibitem[{Xu(2025{\natexlab{b}})}]{Xu2025a}
Xu, Z.-H. 2025{\natexlab{b}}, Research in Astronomy and Astrophysics,
  \dodoi{10.1088/1674-4527/adfeb9}

\bibitem[{{Yu} \& {Tremaine}(2002)}]{yu2002}
{Yu}, Q., \& {Tremaine}, S. 2002, \mnras, 335, 965,
  \dodoi{10.1046/j.1365-8711.2002.05532.x}

\bibitem[{{Zhang} {et~al.}(2025){Zhang}, {Wu}, {Fan}, {Ho}, {Wu}, {Zhang},
  {Lyu}, {Cao}, \& {Wang}}]{Zhang2025}
{Zhang}, C., {Wu}, Q., {Fan}, X., {et~al.} 2025, arXiv e-prints,
  arXiv:2505.12719, \dodoi{10.48550/arXiv.2505.12719}

\end{thebibliography}
\bibliographystyle{aasjournal}

\end{document}